\documentclass[prb,aps,longbibliography,showpacs,groupedaddress,superscriptaddress,twocolumn,toc=flat,nofootinbib]{revtex4-2}

\usepackage{graphicx}
\usepackage{latexsym}
\usepackage{amssymb}
\usepackage{amsmath}
\usepackage{amsfonts}
\usepackage{upgreek}
\usepackage{subfigure}
\usepackage{bm}
\usepackage{bbold}
\usepackage{verbatim}
\usepackage[unicode=true,
 bookmarks=false,
 breaklinks=false,pdfborder={0 0 1},backref=false,colorlinks=true]
 {hyperref}
\hypersetup{
 linkcolor=[rgb]{0,0,1},citecolor=[rgb]{0,0,1},urlcolor=[rgb]{0,0,1}}
\usepackage{multirow}
\usepackage{color}
\usepackage{comment}
\usepackage{placeins}
\usepackage{xcolor}
\usepackage{soul}
\usepackage{tikz}
\usepackage{tabularx}
\usepackage{siunitx}
\usepackage{enumerate}  
\usepackage{xspace} % package for spacing after newly defined commands
\usepackage{ragged2e}

\newcommand{\Harvard}{Department of Physics, Harvard University, Cambridge, MA 02138, USA}
\newcommand{\ITAMP}{ITAMP, Center for Astrophysics $\vert$ Harvard \& Smithsonian, Cambridge, MA 02138, USA}

\let\Re\relax
\DeclareMathOperator{\Re}{Re}
\let\Im\relax
\DeclareMathOperator{\Im}{Im}

\newcommand{\neelfield}{\mathbf{\Omega}}
\newcommand{\NLSM}{NL$\sigma$M\xspace}

\usepackage{braket}

\newcommand{\GDP}[1]{{\color{cyan} GDP: #1}}

\begin{document}

\title{Effective theory of the hidden-order pseudogap phase in a doped antiferromagnet}

\author{Gaia De Paciani}
\thanks{These authors contributed equally to this work.}
%\email{gaia.paciani@lmu.de}
\affiliation{Department of Physics and Arnold Sommerfeld Center for Theoretical Physics (ASC), Ludwig-Maximilians-Universit\"at Munich, Theresienstr. 37, Munich D-80333, Germany}
\affiliation{Munich Center for Quantum Science and Technology (MCQST), Schellingstr. 4, Munich D-80799, Germany}
\author{Gesa Dünnweber}
\thanks{These authors contributed equally to this work.}
\affiliation{Munich Center for Quantum Science and Technology (MCQST), Schellingstr. 4, Munich D-80799, Germany}
\affiliation{Max Planck Institute of Quantum Optics, Hans Kopfermann Str. 1, Garching D-85748, Germany}
\affiliation{Department of Physics, Technical University of Munich, James-Franck-Str. 1, Garching D-85748, Germany}
\author{Johannes Poersch}
\affiliation{Department of Physics and Arnold Sommerfeld Center for Theoretical Physics (ASC), Ludwig-Maximilians-Universit\"at Munich, Theresienstr. 37, Munich D-80333, Germany}
\author{Simon M. Linsel}
\affiliation{Department of Physics and Arnold Sommerfeld Center for Theoretical Physics (ASC), Ludwig-Maximilians-Universit\"at Munich, Theresienstr. 37, Munich D-80333, Germany}
\author{Henning Schl\"omer}
\affiliation{\ITAMP}
\affiliation{\Harvard}
\author{Fabian Grusdt}
\affiliation{Department of Physics and Arnold Sommerfeld Center for Theoretical Physics (ASC), Ludwig-Maximilians-Universit\"at Munich, Theresienstr. 37, Munich D-80333, Germany}
~~\affiliation{Munich Center for Quantum Science and Technology (MCQST), Schellingstr. 4, Munich D-80799, Germany}

\date{\today}
\begin{abstract} 

The microscopic origin of the pseudogap phase constitutes a longstanding puzzle related to the emergence of high-temperature superconductivity in cuprate materials. In this work, we develop an effective model for doped antiferromagnets in terms of fluctuating stripes, or string-like domain walls, which obscure the antiferromagnetic order of the spin background.
The open ends of such domain walls of the Néel order are treated as vortices in the resulting lattice gauge theory.
We numerically evaluate the phase diagram by classical Monte Carlo simulations, using percolation-based geometric order parameters to diagnose hidden Néel order. At high temperatures, we identify a BKT-type crossover in which the domain wall ends become deconfined. We interpret this as the $T^*$ crossover from the hidden order regime to the paramagnetic metal above.  At low temperatures, we identify stripe instabilities. Predictions of our effective model can be tested in ultracold fermion quantum simulators.
 
\end{abstract}

\maketitle

\section{Introduction}\label{sec:intro}
The field of high-temperature superconductivity was established almost 40 years ago when a critical temperature of \(T_c \approx 35\,\mathrm{K}\) was measured in the cuprate compound \(\mathrm{Ba}_x\mathrm{La}_{5-x}\mathrm{Cu}_5\mathrm{O}_{5(3-y)}\) \cite{Bednorz1986}. Ever since, cuprates have been the focal point of theoretical and experimental studies, reaching critical temperatures up to \(T_c \approx 134\,\mathrm{K}\) at ambient pressure \cite{Schilling1993}, well above the boiling temperature of nitrogen at $T \approx \SI{77}{\kelvin}$.

Cuprates feature a complex phase diagram that has yet to be described by a unifying theory \cite{Lee2006, Keimer2015}. Around half filling, the parent compounds are antiferromagnetic Mott insulators. Upon doping, long-range antiferromagnetic (AFM) order is rapidly destroyed~\cite{Lee2006, Xu2009}, superconductivity emerges at low temperature, and a broad pseudogap regime occupies the underdoped part of the phase diagram. The pseudogap is characterized by a partial suppression of low-energy spectral weight~\cite{Shen2005, Kanigel2006, Yang2011, Kurokawa2023} and appears below a doping-dependent temperature scale $T^*$. Its microscopic origin remains one of the central open problems in the theory of cuprates~\cite{Timusk1999, Norman2005, Chowdhury2015}.
\begin{figure}[t!]
    \centering
    \includegraphics[width=0.9\linewidth]{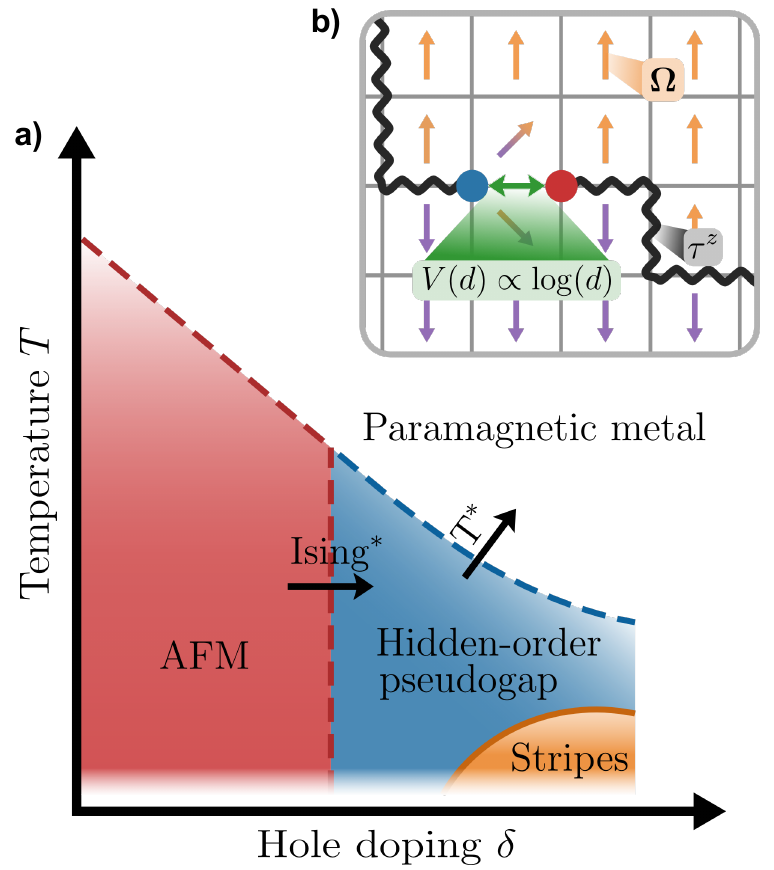}
    \caption{\textbf{Schematic phase diagram of the fluctuating stripes model.}~
    The effective classical model proposed in this work [Eq.~\eqref{eq:Hamiltonian}] describes a system of fluctuating stripes representing the domain walls of an antiferromagnetic background of the spins. Within the phase diagram of this model,\textbf{~a)} we distinguish between the spin-pseudogap regime (AFM), which arises from thermal fluctuations above the two-dimensional Heisenberg antiferromagnet, and a distinct \textit{hidden-order} pseudogap phase that originates from the proliferation of spatially-fluctuating stripe-like domain walls. The region that we identify as hidden-order pseudogap is characterized by hidden-AFM quasi-long-range order of the spin background.\textbf{~b)} The inset illustrates the effect of the domain walls, represented by $\tau^z$, on the underlying Néel order parameter~$\Omega$: the ends of the domain walls behave as vortices and the interaction between open ends scales logarithmically with their distance. The assumption made for the consistency of this effective model is that the correlation length $\xi^*$ of the spin background must be much larger than the distance of the open ends of the strings, i.e. $\xi^* \gg d$. This ensures that the spins rotate in a plane in spin space, yielding an effective $U(1)$ order parameter~(see App.~\ref{sec:methods_VortexIntMotivation}).} 
    \label{fig:Cuprate_Phase_Diagram}
\end{figure}

Several scenarios have been proposed for the pseudogap. In one class of ideas, it is viewed as a precursor of superconductivity, associated with incoherent or preformed Cooper pairs~\cite{Wang2006, Yang2008, Kanigel2008, niu2024e}. In another, it is tied to competing or intertwined symmetry-breaking orders, such as charge\nobreakdash-density waves or spin stripes~\cite{Millis2007, Dimov2008, Norman2010, Yao2011}. A third class of proposals interprets the pseudogap as a distinct metallic phase~\cite{Chowdhury2014, Chowdhury2015, Punk2015}, possibly involving fractionalization~\cite{Senthil2003} or an orthogonal-metal-like state~\cite{Ruegg2010, Nandkishore2012, Gazit2020}. These viewpoints are not mutually exclusive at the phenomenological level. Yet they suggest rather distinct microscopic mechanisms, leading to significantly different interpretations of the $T^*$ line in the phase diagram.

Here, we analyze a proposal that builds on the close relationship among doping, magnetism, and stripe physics. Although static long-range AFM order disappears already at small hole concentration, substantial short-range AFM correlations and spin-wave-like excitations persist well beyond the underdoped regime \cite{Tranquada1996, Motoyama2007}. At low temperatures, many cuprates also exhibit spin and charge stripe order, or at least strong stripe-like correlations~\cite{Emery1999, Kivelson2003, Vojta2009, Tranquada1995, Tranquada1996, Abbamonte2005}. These observations suggest that doped holes do not simply melt the AFM background uniformly. Instead, they may reorganize it through fluctuating antiferromagnetic domain walls of the antiferromagnetic Néel order $\Omega$, as illustrated in Fig.~\ref{fig:Cuprate_Phase_Diagram}.

Recently, a geometric scenario was proposed in which such spatially fluctuating string-like domain walls provide the microscopic origin of the doped pseudogap phase~\cite{Schloemer2025}. In this picture, the spin background retains \textit{hidden-AFM order}. The order is not visible directly through two-point correlations because fluctuating domain walls interchange the sublattice parity in spatial regions of the sample. However, it can be restored in a squeezed reference frame in which the domain-wall structure is removed. One theoretical scenario is that the resulting state realizes a geometric orthogonal metal: fermionic quasiparticles remain well defined in the hidden-AFM background, while their physical-electron spectral weight is suppressed by the fluctuating string network~\cite{Schloemer2025}.

This mechanism naturally connects three components of the cuprate problem: AFM correlations, stripes, and the pseudogap. Closed fluctuating domain walls can obscure conventional AFM two-point correlations without fully destroying the underlying spin order. Their proliferation gives rise to hidden order by concealing the existing AFM correlations, while their topology suggests a natural language in terms of an emergent \(\mathbb{Z}_2\) gauge structure~\cite{Wilke2025}. In this framework, the transition from long-range AFM order to hidden-AFM order is non-local and is expected to be diagnosed most directly by geometric observables, rather than by conventional local order parameters or two-point correlations.

In this work, we investigate the finite-temperature fate of this hidden-order scenario for the doped pseudogap phase. We formulate an effective classical string model for doped antiferromagnets in which Néel domain walls are represented by Ising variables \(\tau^z_{\langle i,j\rangle}\) living on the links of a square lattice. Closed strings describe fluctuating AFM domain walls. Open string ends are allowed and are interpreted as topological defects of the hidden-AFM background. These defects carry vorticities \(u_i=\pm 1/2\) and interact through an effective logarithmic potential, as expected for vortex-like excitations in two dimensions.

The central result of our analysis is that -- within the scenario we propose -- the pseudogap temperature $T^*$ can be understood as a confinement-deconfinement transition of these open string ends. At low temperature, topological charges remain confined in bound vortex-antivortex pairs, and the string network can support hidden-AFM order. At elevated temperature, these pairs proliferate and unbind. We argue that this destroys the hidden-order construction and drives a transition into a paramagnetic metal or Fermi-liquid-like regime, see Fig.~\ref{fig:Cuprate_Phase_Diagram}.

We study the phase diagram of the effective model using classical Markov chain Monte Carlo simulations. To characterize the geometry of the fluctuating strings, we use percolation order parameters (POPs), which were recently introduced as non-local diagnostics of confinement in \(\mathbb{Z}_2\) lattice gauge theories \cite{Linsel2024, Duennweber2025, Linsel2026}. In the present setting, POPs detect whether the domain-wall network forms system-spanning clusters. This allows us to distinguish a low-doping regime with non-percolating loops and where we expect to have quasi-long-range AFM order from an intermediate regime with percolating fluctuating stripes and hidden-AFM order, in the regime of confined string-endpoints.

To diagnose the confinement of open string ends, we further introduce a $U(1)$ correlation function constructed from the phase field associated with the vortex defects. Its long-distance behavior distinguishes algebraic correlations from exponential decay. We find a transition from power-law to exponential behavior upon increasing temperature, indicating a Berezinskii--Kosterlitz--Thouless (BKT) type crossover. We associate this crossover of hidden-AFM order with the experimentally observed $T^*$ scale.

Our results support the following physical picture, summarized in Fig.~\ref{fig:Cuprate_Phase_Diagram}. At low doping, the system is dominated by short, non-percolating domain-wall loops and is continuously connected to the AFM regime. At intermediate doping and low temperature, fluctuating domain walls percolate while their open ends remain confined. This regime realizes a hidden-AFM, doped-pseudogap phase. At higher temperatures, vortex-antivortex pairs at the ends of domain walls deconfine through a BKT-like mechanism, destroying the hidden-order structure and fully restoring the $SU(2)$ symmetry. In this way, the pseudogap boundary at $T^*$ is associated not with the onset of a local order parameter, but with a topological change in the fluctuating domain-wall ensemble.
As an important consequence, the scenario discussed in this article predicts two distinct types of pseudogap regimes, connected to one another. At zero and ultra-low doping, we expect the well-known AFM-pseudogap, associated with the onset of extended AFM order in two dimensions and marked by AFM correlation length $\xi \gg a$~\cite{Chalopin2025, Kendrick2025}, where $a=1$ is the lattice constant. In Fig.~\ref{fig:Cuprate_Phase_Diagram} we denote this simply as ``AFM''. At higher doping, we propose the distinct hidden-order pseudogap to emerge from the AFM-pseudogap through an Ising$^*$ transition, i.e., without a local order parameter. This phase is characterized by a significantly shorter AFM correlation length $\xi$ due to the proliferation of AFM domain walls, whilst hidden AFM correlations are associated with a second length scale $\xi^* \gg \xi$. As we show below, the Ising$^*$ transition manifests in pronounced maxima of compressibility, producing a Widom line, as theoretically predicted~\cite{Sordi2012} and recently observed in cold atom experiments~\cite{Kendrick2025}. 

The remainder of the paper is organized as follows. In Sec.~\ref{sec:eff_string_model}, we introduce the effective string model and motivate its relation to fluctuating AFM domain walls. In Sec.~\ref{sec:t-star-transition}, we present the Monte Carlo results, including the percolation analysis, vortex density, and \(U(1)\)-correlator diagnostics of the BKT-like transition from hidden-order to paramagnetic metal. In Sec.~\ref{sec:discussion}, we discuss the resulting interpretation of the effective model and the obtained phase diagrams and summarize the implications for the $T^*$ line.

\section{Effective String Model}\label{sec:eff_string_model}
Our starting point is the experimentally and numerically motivated observation that doped holes can bind into stripes that generate domain walls in the Néel order of the AFM background~\cite{Tranquada1995, Tranquada1996, Emery1999, Vojta2009}. At zero temperature, we assume that this binding is robust, i.e. that the energy required to remove a hole from a stripe ($E_{\mathrm{bind}}$) is much larger than the interaction energy between different stripes ($E_{\mathrm{stripe\text{-}stripe}}$). The latter controls the spatial ordering of the stripes relative to each other. Due to this hierarchy of the energy scales, heating the system can dissolve the spatial ordering of ground state stripes -- restoring spontaneously broken lattice symmetry -- without unbinding holes from N\'eel domain walls. Instead, it thermally disorders an ensemble of intact but fluctuating stripe segments. The model introduced below describes this particular melting scenario, retaining the geometry and topology of the stripes while effectively integrating out the microscopic motion of the holes bound to them.
It serves as an effective classical statistical model for the geometric sector that we propose to be central to the doped pseudogap regime.

In this formulation, the microscopic electronic and spin degrees of freedom are replaced by a small number of effective parameters. The remaining variables retain the information needed to distinguish the relevant geometric regimes: the position of domain-wall strings, the location of their open ends, and the long-distance interaction between these endpoints.

Specifically, doped holes are assumed to
bind into line-like $\pi$ domain walls: crossing them changes the sublattice parity and produces a $\pi$ phase-shift of the local Néel order parameter. We therefore coarse-grain each hole-rich stripe to a one-dimensional string-like domain wall. The microscopic stripe width and the internal arrangement of holes along it are not resolved; their effects enter only through effective couplings. 

We represent these domain walls as Ising variables on the nearest-neighbour links of a square lattice $\{ \tau^z_{\langle \mathbf{i}, \mathbf{j} \rangle } \}$, with $\tau^z_{\langle \mathbf{i}, \mathbf{j} \rangle } = -1$ if there is a domain wall across the link $\langle \mathbf{i}, \mathbf{j} \rangle$; and $\tau^z_{\langle \mathbf{i}, \mathbf{j} \rangle } = +1$ otherwise, see Fig.~\ref{fig:Cuprate_Phase_Diagram}b).
The hole-doping enters the effective theory through the total number of strings. We quantify it through the mean fraction of occupied links,
\begin{equation}\label{eq:density}
    \delta=\nu\left\langle\frac{1-\tau^z}{2}\right\rangle.
\end{equation}
Here $\nu$ is a phenomenological conversion factor that must represent both the filling of holes along a stripe and the geometric conversion from link density to hole density per lattice site. We set $\nu=1$ in the following. Thus $\delta$ is an \emph{effective doping}, not a direct representation of the microscopic hole concentration. With this convention, a maximally disordered link ensemble has $\delta=1/2$. The coupling $h$ introduced below controls the occupied-link density and therefore acts as a chemical potential for the stripes; varying $h$ changes the effective doping indirectly through Eq.~\eqref{eq:density}.

\subsubsection*{A. Closed loop case.} \label{subsec: closed_loops}
The domain-wall representation explains how fluctuating stripes can hide antiferromagnetism without destroying the information about the long-range Néel order of the background. The Néel orientation between two sites can be compared after accounting for every $\pi$ shift produced by the domain walls crossed along a connecting path. If the strings form closed loops, this reconstruction is path independent within the topological sector: crossing a domain wall changes the apparent sublattice parity in real space, but this change can be undone in the reconstructed, or squeezed, reference frame~\cite{Wilke2025, Schloemer2025}. Ordinary two-point spin correlations may therefore be short-ranged even though antiferromagnetic order remains visible after the domain-wall geometry has been removed. We call this reconstructed order \emph{hidden} AFM order. 

We introduce the minimal Hamiltonian for a system with closed loops:
\begin{equation}\label{eq: minimal_closed_loop_model}
    \mathcal{H} = \lim_{\mu \rightarrow -\infty} \big[ - h \sum_{\langle \mathbf{i},\mathbf{j} \rangle } \tau^z_{\langle \mathbf{i},\mathbf{j} \rangle} 
    - \mu \sum_{\mathbf{i}} n_{\mathbf{i}} \big].
 \end{equation}
The first term in Eq.~\eqref{eq: minimal_closed_loop_model} controls the string tension. For $h>0$, configurations with occupied links ($\tau^z_{\langle \mathbf{i},\mathbf{j} \rangle} = -1 $) are penalized relative to the empty lattice; Increasing the temperature or decreasing the effective string tension promotes longer fluctuating strings.
On the lattice, the endpoints of open strings are determined by the parity of the number of occupied links touching a site. 
We define
\begin{equation} \label{eq:top_charges}
n_\mathbf{i} \equiv \frac{1}{2} \left( 1 - \prod_{\mathbf{j}: \langle \mathbf{i}, \mathbf{j} \rangle}  \tau^z_{\langle \mathbf{i}, \mathbf{j} \rangle } \right)
\end{equation}
such that $n_\mathbf{i} = 1$ occur at exactly those sites $\mathbf{i}$ with an odd number of incoming strings (otherwise $n_\mathbf{i} = 0$).
The model of Eq.~\eqref{eq: minimal_closed_loop_model} is dual to the $2D$ Ising model~\cite{Peierls1936, Wegner1971}
\begin{equation}\label{eq: hamiltonian_ising}
    \mathcal{H} = - h \sum_{\langle \mathbf{i},\mathbf{j} \rangle } \sigma^z_{\mathbf{i}}\sigma^z_{\mathbf{j}} + \mathrm{const}
 \end{equation}
with $\sigma^z_{\mathbf{i}}$ being classical Ising spins living on the sites of the lattice. Given this duality mapping, tuning $h$ in Eq.~\ref{eq: minimal_closed_loop_model} drives an Ising$^*$ transition from a non-percolating to a percolating regime~\cite{Linsel2024, Schloemer2025}.
Increasing doping increases the available domain-wall strings. Short, non-percolating loops weakly perturb the antiferromagnet, while a percolating network can strongly conceal real-space Néel order, still preserving its hidden reconstruction.
This picture can, in principle, justify spin-wave-like excitations in squeezed space and accommodate magnetic-polaron quasiparticles. Once quantum and topological fluctuations are restored, it may also support a small Fermi surface and geometric orthogonal-metal phenomenology without requiring static stripe order~\cite{Schloemer2025}. The purpose of the present work is to isolate the minimal classical geometric sector and ask when the hidden-order reconstruction ceases to be well defined.
\subsubsection*{B. General case: open domain walls.}\label{subsec: open_domain_walls}
We introduce open domain walls, associated with defects of the reconstructed Néel field. We focus on understanding whether such endpoints remain confined in pairs or proliferate and deconfine. Confined vortex-antivortex pairs of the Néel order produce only local ambiguities and leave the hidden-order construction meaningful at long distances, while unbound endpoints destroy its global consistency~\cite{DePaciani2026}.
Each open end ($n_\mathbf{i}=1$) carries a vorticity $u_\mathbf{i} = \pm 1/2$, which denotes the two possible windings of the Néel field around those endpoints.
To motivate the interaction between these defects, we start from the non-linear $\sigma$-model, which is a well established description of antiferromagnets near half filling~\cite{Chakravarty1989, Fradkin2013}. Its low-energy field is a unit Néel vector $\neelfield(\mathbf{x})$, with $|\neelfield|=1$. A stripe corresponds to a narrow spatial region, of order the lattice spacing $a$, across which $\neelfield$ rotates to $-\neelfield$. This imposed $\pi$ rotation contributes a local energy proportional to the stripe length. Near an endpoint, the Néel field must also be distorted in the surrounding two-dimensional region.

In reducing this distortion to a Coulomb-gas interaction, we make the assumption that the separation $d$ of the relevant endpoint pair is smaller than the hidden-antiferromagnetic correlation length, $a\ll d\ll\xi^*$, so that the background is ordered on the scale of the pair. This justifies a picture in which the spin-distortion, which minimizes the energy, is locally coplanar, yielding an effective $U(1)$ order parameter. Under these assumptions, an endpoint is a half-vortex with vorticity $u_{\mathbf{i}}=\pm1/2$, corresponding to a winding $2\pi u_{\mathbf{i}}=\pm\pi$. Its long-wavelength elastic energy gives the two-dimensional Coulomb interaction $-u_{\mathbf{i}}u_{\mathbf{j}}\log(r_{\mathbf{i}\mathbf{j}}/a)$ derived in Appendix~\ref{sec:methods_VortexIntMotivation}, with $r_{\mathbf{i}\mathbf{j}}$ the distance between the two vortices.

The minimal Hamiltonian we will analyze in the remainder of this article thus reads:
\begin{equation} \label{eq:Hamiltonian}
    \mathcal{H} = - h \sum_{\langle \mathbf{i},\mathbf{j} \rangle } \tau^z_{\langle \mathbf{i},\mathbf{j} \rangle} 
    - \mu \sum_{\mathbf{i}} n_{\mathbf{i}} 
    - \frac{1}{2}M \sum_{\mathbf{i} \neq \mathbf{j}} n_{\mathbf{i}}n_{\mathbf{j}}u_{\mathbf{i}}u_{\mathbf{j}}\mathrm{log}\vert \mathbf{i}-\mathbf{j} \vert  .
\end{equation}
The corresponding statistical physics problem considers a thermal state, $\rho(\{ \tau^z_{\langle \mathbf{i},\mathbf{j} \rangle}, u_{\mathbf{j}}\}) = e^{-\beta\mathcal{H}}/Z$, with string degrees of freedom $\tau^z_{\langle \mathbf{i},\mathbf{j} \rangle}$ and vorticities $u_{\mathbf{j}}=\pm 1/2$ on the open ends of the strings, where $n_{\mathbf{j}}= 1$.
We treat the number of string-endpoints, $N_V = \sum_i n_i$, grand-canonically, i.e. $N$ is controlled by $\mu$. 
%We take $M<0$, so that isolated vortices have a positive energy cost. 
In the limit $\mu\rightarrow -\infty$, vortices are completely suppressed and the model reduces to a closed-loop theory.
The last term of  Eq.~\eqref{eq:Hamiltonian} is the logarithmic interaction between vortices inherited from the long-wavelength deformation of the Néel field. For $M>0$, defects of opposite vorticity attract, while defects of equal vorticity repel. 
\begin{figure}[t!]
    \centering
    \includegraphics[width=\linewidth]{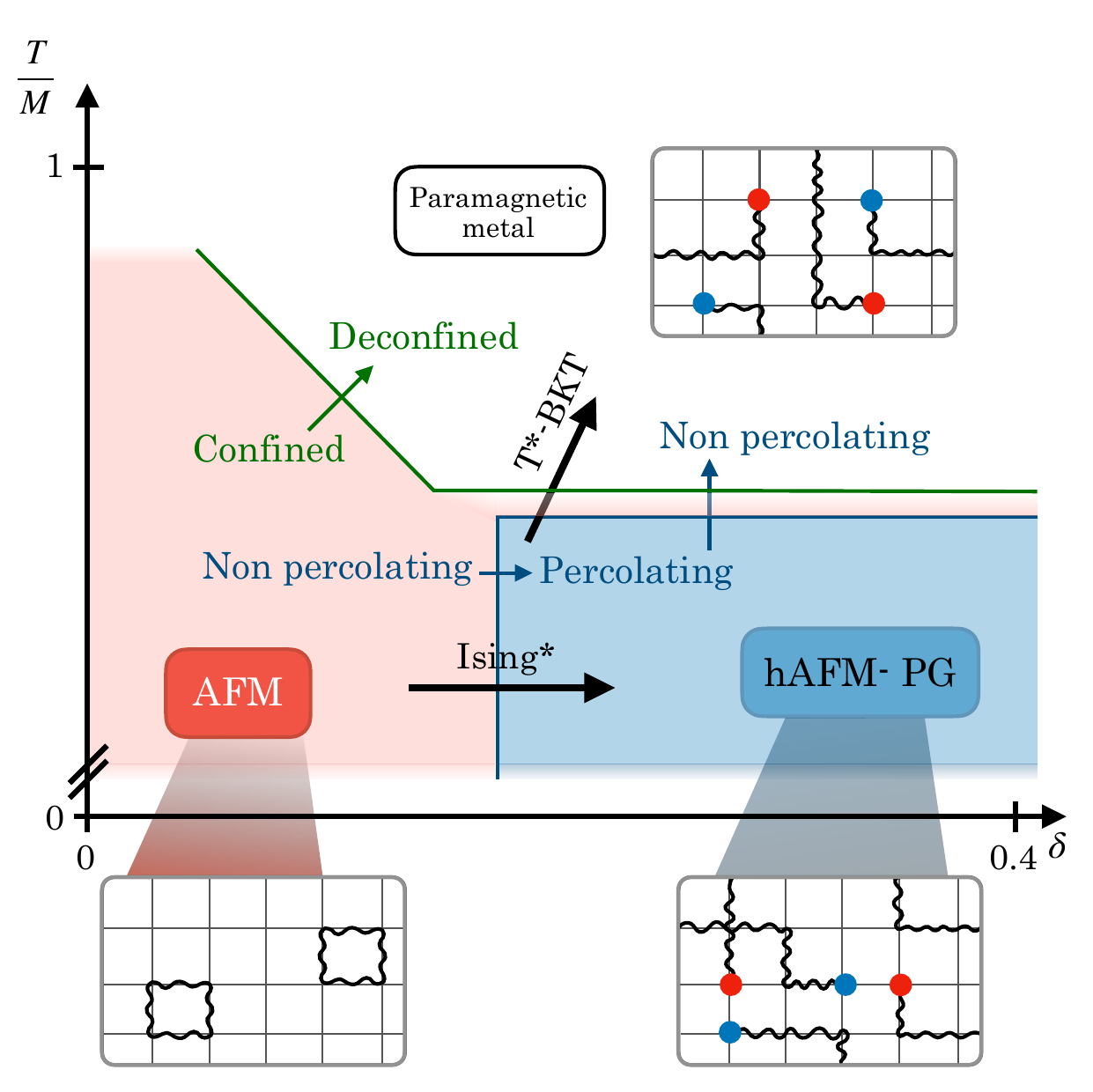}
    \caption{\textbf{Qualitative phase diagram of the effective string model}, Eq.~\eqref{eq:Hamiltonian}. Temperature $T/M$ is plotted over the effective doping $\delta$, Eq~\eqref{eq:density}. We find that the model undergoes a percolation (Ising$^*$) transition at a finite effective doping level. This transition marks the evolution from the AFM pseudogap phase to the hidden-AFM regime. Additionally, we find a BKT phase transition that we identify as the $T^*$ transition outside of the pseudogap regime. Insets show the representative snapshots in the different regimes of the studied model.} 
    \label{fig:Model_Interpretation}
\end{figure}

The logarithmic vortex interaction is the key reason to expect a Berezinskii-Kosterlitz-Thouless-type transition or crossover~\cite{Kosterlitz1973}. At low temperature and low vortex density $\rho=N/L^2$, the logarithmic attraction confines opposite-vorticity endpoints into bound pairs. In this regime, open strings occur only as virtual defects on a background dominated by fluctuating closed domain walls. At sufficiently high temperature, entropy can overcome the logarithmic attraction and vortex pairs unbind. As we show in the next section, the resulting deconfinement of vortices destroys the hidden-AFM structure and provides a mechanism for the $T^*$ transition.

The simplicity of Eq.~\eqref{eq:Hamiltonian} also means that it allows configurations that may be expected to be strongly suppressed in a more realistic description. In particular, the minimal model does not distinguish between horizontal and vertical stripe tendencies, does not prohibit domain-wall crossings, and does not include an energetic preference for globally aligned stripes. These ingredients can be added through further effective interactions. In Appendix~\ref{sec:methods_stripes} we consider such extensions, including terms that penalize crossings and favor locally aligned domain-wall segments while preserving the microscopic $C_4$ symmetry of the lattice. The qualitative confinement physics associated with the logarithmically interacting vortices, however, is already present in the minimal model. For this reason, Eq.~\eqref{eq:Hamiltonian} provides the simplest setting in which to test whether a classical ensemble of fluctuating domain walls can support a hidden-AFM pseudogap regime and a finite-temperature deconfinement transition.

\section{$\mathbf{T^*}$ transition} \label{sec:t-star-transition}
\begin{figure*}[t!]
    \includegraphics[width=1.\linewidth]{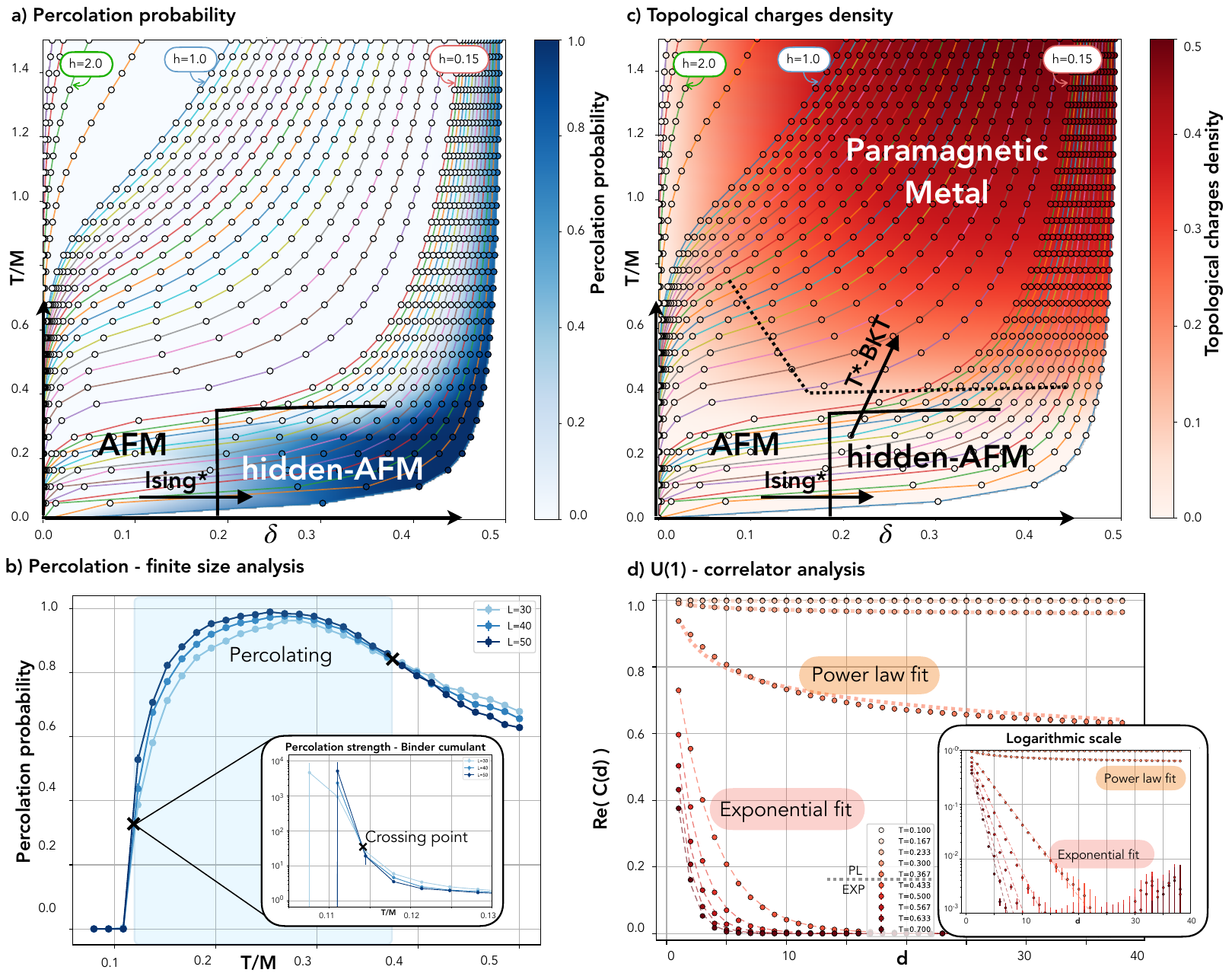} 
    \begin{minipage}{\textwidth}
    \caption{ \textbf{Monte Carlo analysis of the effective model in Eq.~\eqref{eq:Hamiltonian}.}~\textbf{a)} We plot the percolation probability at different points in the temperature vs effective doping phase diagram obtained from Monte Carlo simulations with system size $L=50$. Solid lines connect data points with a fixed value of $h$ (a selection indicated on top). To extrapolate the continuous color background on the entire doping-temperature space, a Delaunay interpolation is performed. We identify the percolating region with the hidden-AFM order, whereas the non-percolating region at low temperatures is associated with long-range AFM order.~\textbf {b)} To extrapolate and verify the percolating regimes, we perform a finite-size analysis on the percolation probability and check the percolation strength binder cumulant crossing point (inset). We determine the regimes where the system enters and exits the percolation regime, for fixed $h=0.05$.~\textbf{c)}~The vortex density $\langle n_i \rangle$ is plotted in the temperature vs. effective doping phase diagram. The dotted line shows the estimated location of the BKT-transition, which is determined through the analysis of the $U(1)$~order parameter.~\textbf{d)}~Plots of the average $U(1)$ correlation $C(d)$ at distance $d$: for each temperature step at fixed $h=0.2$ we examine the behavior of the correlator to roughly estimate whether, and in which temperature regime, we detect the BKT-characteristic power-law to exponential crossover. The full points are the results of the Monte Carlo analysis, the dashed line represents the best exponential fit obtained while the dotted lines represent the best power law fit obtained. At roughly $T\geq0.43$ the best fit corresponds to exponential decay, whereas at $T \leq 0.367$, the best fit yields power law decay.}
    \label{fig:Monte_Carlo_results}
    \end{minipage}
\end{figure*}
We now study the phase diagram obtained from the Hamiltonian of Eq.~\eqref{eq:Hamiltonian} through a classical Monte Carlo analysis. Our results are schematically summarized in the sketch of the phase diagram in Fig.~\ref{fig:Model_Interpretation}. 

In order to diagnose the transition from AFM to hidden-AFM, which is non-local in terms of the string variables $\tau^z$, we use the \textit{percolation probability} $\Pi$, defined as the probability that there exists a connected cluster of strings that spans the entire lattice~\cite{Linsel2024, Duennweber2025, Linsel2026}~(see App.~\ref{supp_sec: numerical details} for more details). In Fig.~\ref{fig:Monte_Carlo_results}a, we plot the percolation probability. 
We infer the effective doping from the average string filling of the snapshots, and we plot the result as a function of the effective doping $\delta$, see Eq.~\eqref{eq:density}.

The results of Fig.~\ref{fig:Monte_Carlo_results} are obtained from independent classical Markov Chain Monte Carlo simulations with fixed parameters $M = -\mu = 1.0$ and varying~$h$. The coloured curved lines represent different values of $h$, going from $h \ll M$ to $h = \mathcal{O}(M)$ (i.e. $h = 0.01 M$ to $h=3.0 M$). Each point corresponds to one simulation at a given temperature and $h$, and the hole density is determined from the average string density according to Eq.~\eqref{eq:density}. Further, we employ finite-size analysis to verify the reliability of the percolation results; an example of this is given in Fig.~\ref{fig:Monte_Carlo_results}b, where we show the percolation probability at different system sizes and the binder cumulant crossing of the percolation strength for $h=0.05$. We adopt mixed boundary conditions, meaning that the strings are evolved with periodic boundary conditions, while the interaction between vortices is evaluated using the direct Euclidean distance rather than the periodic metric (see App.~\ref{supp_sec: numerical details} for further details). 

In Fig.~\ref{fig:Monte_Carlo_results}c we plot the density of the vortices in a similar manner as the percolation transition of Fig.~\ref{fig:Monte_Carlo_results}b.
The plot shows that increasing temperature leads to progressively increasing vortex density.
We define the \textit{$\mathit{U(1)}$ correlation parameter}:  
\begin{align}   \label{eq:U1_correlator}
    C(|\mathbf{x} - \mathbf{x}'|) &= \Re \left\langle 
        \mathrm{e}^{i\{\phi(\mathbf{x}) -\phi(\mathbf{x}')\}}
    \right\rangle  
\nonumber \\
    \text{with } 
    \phi(\mathbf{x}) &= \sum_{\mathbf{j}: n_{\mathbf{j}} = 1} -u_{\mathbf{j}} 
    \Im\{\log ( \mathbf{x} - \mathbf{x}_{\mathbf{j}}) \},
\end{align}
via the average $U(1)$ phase acquired by each open end from the interaction with the others (see Sec.~\ref{sec:methods_U1op}), where $\mathbf{x} = x + iy$ and $x$ and $y$ are the coordinates of the vertices of the lattice. 
We use the latter to detect a phase transition from a confined phase to a deconfined phase; Specifically, we observe a power law to exponential change of behavior in the plots of the correlations, see Fig.~\ref{fig:Monte_Carlo_results}d, as the temperature increases.

The Monte Carlo results confirm the picture discussed in Sec.~\ref{sec:intro} and~\ref{sec:eff_string_model} and provide a direct real-space characterization of the different regimes of the effective theory. Fig.~\ref{fig:Monte_Carlo_results} provides a numerical confirmation of the geometric and topological view of the phase diagram of Fig.~\ref{fig:Model_Interpretation}. At low temperature, Fig.~\ref{fig:Monte_Carlo_results}a shows a clear change with increasing doping: the percolation probability is close to zero in the low-density region labeled \textit{AFM}, where we expect a regime with long-range AFM order. The percolation probability becomes large once the strings join into a system-spanning network, rendering two-point spin-spin correlations
in the microscopic model short-ranged. In our effective model, percolation marks the point at which fluctuating domain walls cease to be dilute local distortions and instead obscure the Néel pattern on the scale of the whole sample. The fixed-$h$ plot in Fig.~\ref{fig:Monte_Carlo_results}b shows this in more detail: the rapid rise of $\Pi$ near $T/M\simeq0.12$ becomes sharper with increasing $L$, and the crossing of the corresponding percolation strength Binder ratios near the same temperature supports a genuine transition. 
Percolation alone, however, does not establish whether hidden order survives, because the reconstruction of the N\'eel background also requires the domain-wall endpoints to remain confined to stabilize the hidden-AFM order. Fig.~\ref{fig:Monte_Carlo_results}c brings this second piece of information. The vortex density is low throughout the low-temperature AFM and hidden-AFM regions, but clearly increases upon heating; the dotted line therefore tracks the onset of the regime with proliferating vortices. Finally, Fig.~\ref{fig:Monte_Carlo_results}d shows in detail a confinement-check through the long-distance behaviour of $C(d)$ for the example $h=0.2
$. At lower temperatures, the correlations are algebraic, as expected for bound vortex--antivortex pairs. Beyond a critical $T$ ($T/M\geq0.433$ in this specific instance) they are better described by exponential decay. 
Overall, this analysis allows us to distinguish two logically separate changes of order: the percolation transition converts the low-density AFM regime into a hidden-AFM regime of percolating domain walls, while the high-temperature unbinding of the vortices destroys the hidden order and leads to the regime that we identified with the paramagnetic metallic phase. The latter cannot be captured by our effective model, but is characterized by the complete
restoration of $SU(2)$ symmetry indicated via vortex-artivortex deconfinement.

Since the hole density is inferred from the average string density through Eq.~\eqref{eq:density}, the parameter $h$ acts in the effective description as the chemical potential of the relevant $\tau$-variables. Accordingly, we define the charge density $n=1-\delta$ and compute compressibility as
\begin{equation} \label{eq:compressibility}
    \kappa = \frac{1}{n^2}\frac{\partial n}{\partial h},
\end{equation}
which directly probes the reorganization of the string ensemble. In particular, a strong enhancement of $\kappa$ is expected when the charge correlations change rapidly, namely when the system crosses from a regime of short, non-percolating loops to one with extended, percolating fluctuating stripes. In the dual formulation of our model obtained in the closed-loops limit, see Sec.~\ref{subsec: closed_loops}, this transition is manifested by the Ising$^*$ structure of the fluctuating-loop sector.

\begin{figure}[t!]
    \centering
    \includegraphics[width=1.\linewidth]{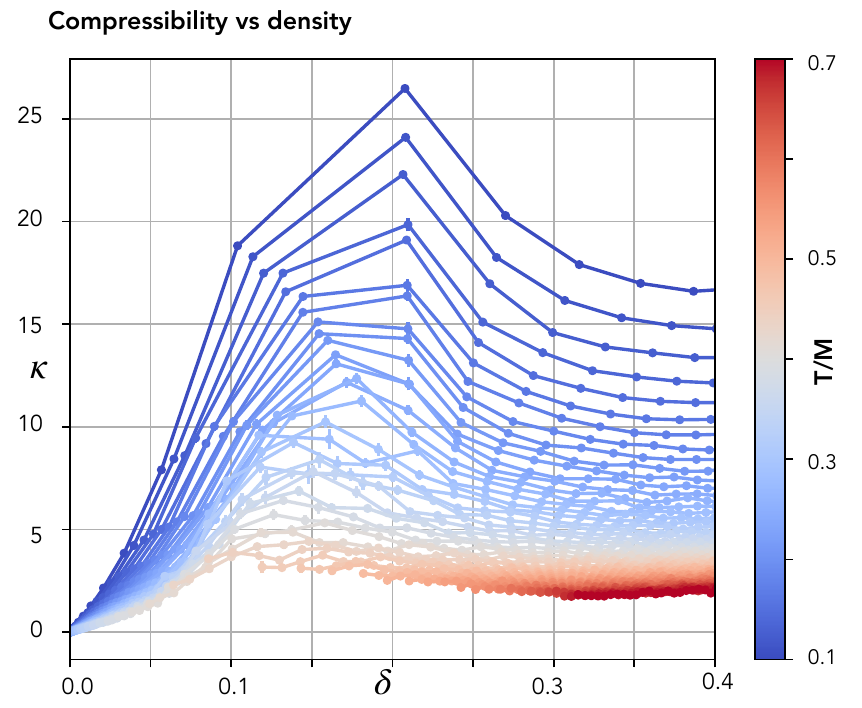}
    \caption{\textbf{Compressibility peak from the percolation transition.} We show the behavior of the compressibility, calculated in the effective model using Eq~\eqref{eq:compressibility}, for different temperatures and effective hole-doping $\delta$. In the low- T regime, where open strings are confined, a pronounced maximum of the compressibility (Widom line) is revealed around the percolation transition. For higher temperatures, where open strings deconfine, the maximum, becomes less pronounced while the compressibility flattens out and decreases.}  \label{fig:Compressibility}  
\end{figure}

In Fig.~\ref{fig:Compressibility}, we plot $\kappa$ as a function of~$\delta$.
The response changes qualitatively across the temperature range. At high temperature, $\kappa(\delta)$ forms a broad, nearly featureless background, indicating that the string density evolves smoothly with $h$ and that no selected density controls the reorganization of the ensemble.
Upon lowering the temperature, a pronounced maximum develops around $\delta\simeq0.15$--$0.22$ and sharpens strongly, reaching its largest amplitude at the lowest simulated temperatures. Its position coincides, within the resolution of the parameter scan, with the low-temperature percolation boundary between the AFM and hidden-AFM regions in Fig.~\ref{fig:Monte_Carlo_results}a. Moreover, the peak is prominent only in the low-temperature window in which Fig.~\ref{fig:Monte_Carlo_results}c,d indicates confined vortices; it broadens and is largely washed out on approaching the BKT unbinding scale $T/M\simeq0.4$. The compressibility maximum therefore tracks the same temperature and density scales as the geometric reorganization of the strings, whereas the high-temperature deconfined regime has no comparably sharp compressibility signature, see App.~\ref{supp: compressibility} for further discussion on the compressibility/Widom line.

Within the introduced framework, the phase diagram can be separated into three main regimes:

a) \textit{Non-percolating confined regime} (AFM / spin-pseudogap regime):
At low string density, which in our interpretation corresponds to low hole concentration $\delta$, the system is dominated by short closed loops. The snapshots obtained from the simulations display neither strings that span the entire system nor a proliferation of isolated vortices, meaning that $\langle \Pi \rangle \approx 0$ and that topological charges remain bound in pairs of opposite vorticity. The AFM background is hence only weakly perturbed in real space. In our data, local string fluctuations, such as elementary plaquette deformations and short-lived bound pairs of vortices, appear but remain dilute and do not modify the large-scale string pattern of the system. 
We associate this low-doping regime with an AFM phase at low temperature. At elevated temperature, the spin correlations are already short-ranged (as explained by the non-linear sigma-model) and decay exponentially. Prior works have shown that the spin susceptibility exhibits a maximum at some temperature $T^{\ast}$, so that this same part of the phase diagram may be viewed as a \textit{spin-pseudogap} regime~\cite{Tanaka2025}. In this sense, we speculate that the low-doping pseudogap phenomenology is tied primarily to the spin sector rather than stripe percolation.

b) \textit{hidden-AFM pseudogap regime}:
Upon increasing the string density, the closed loops grow and eventually form system-sized clusters. The characteristic feature of this regime is the coexistence of percolation of the string network with confinement of vortices. The latter enables the reconstruction of the background Néel order across pairs of open strings.
In real space, the fluctuating domain walls separate the lattice into extended regions related by a $\pi$ shift of the Néel order $\neelfield(\mathbf{x})$~\cite{Schloemer2025}. We identify this regime with a hidden-AFM phase, or the charge-pseudogap phase.
The reorganization of the stripe order provides a natural interpretation of the pronounced step in the density, or equivalently, the enhancement of the compressibility, between the two regimes.  An approximate Widom line can be extracted from the position of compressibility maxima and should be understood as lying approximately at the crossover between AFM and hidden-AFM, although the extracted line does not necessarily coincide exactly with the sharp transition line of the corresponding dual Ising description (see Appendix~\ref{supp: compressibility}).

c) \textit{Deconfined high-temperature regime}:
At higher temperatures, the density of vortices increases substantially, and the confined description can not be applied anymore. In this part of the phase diagram, the system no longer consists primarily of fluctuating closed domain walls, but instead open strings proliferate, and the endpoints of these deconfine. The $U(1)$ correlator of Eq.~\eqref{eq:U1_correlator} changes its long-distance behavior from algebraic to exponential decay, consistent with a BKT-like transition.
The nature of the fluctuations destroying the charge pseudogap is distinct from the low-$T$ Ising$^*$ transition. For this reason, the high-temperature destruction of the charge-pseudogap regime does not lead to a sharp compressibility maximum analogous to the low-temperature case. Instead, the compressibility tends to flatten at large temperature and high doping, consistent with a crossover toward a more conventional Fermi-liquid-like regime, see Fig~\ref{fig:Compressibility}.

The connection of the effective model to the microscopic cuprates picture breaks down in this regime and the high-temperature results must be interpreted with care. In deriving the logarithmic interaction, we assumed that a relevant vortex pair of size $d$ is embedded in a coherent hidden-AFM background, so that $d\ll\xi^*$. On approaching the BKT transition, vortex pairs occur on progressively larger scales; above it, free vortices proliferate and there is no finite pair size $d$ that controls the long-distance physics. At the same time, thermal fluctuations reduce $\xi^*$. Once the $d$ scale becomes comparable to or larger than $\xi^*$, the locally coplanar $U(1)$ description used in Eq.~\eqref{eq:Hamiltonian} stops being justified.
We therefore expect no form of (quasi) long-range AFM to be possibly sustainable and that, therefore, a paramagnetic metal will be realized above $T^*$.

\section{Discussion \& Conclusion} \label{sec:discussion}
In conclusion, we formulated an effective theory of fluctuating stripes for cuprates, including the possibility of open ends. 
Specifically, the present construction should be viewed as an effective theory of the strong-coupling regime, in which the confinement and fluctuation of domain walls provide the leading organizing principle of the spin and charge degrees of freedom in a doped AFM Mott insulator. Within our interpretation, the low-doping AFM region continuously connects to a \textit{spin-pseudogap} regime, while the higher-doping hidden-AFM region realizes a distinct \textit{charge-pseudogap} regime associated with fluctuating and percolating stripes. We found that a percolation transition separates these two regimes of the phase diagram.
The two can be meaningfully distinguished only below the temperature scale $T^*$ at which we assume the antiferromagnetic correlations remain sufficiently robust to allow the entire hidden-order/squeezed-space construction to be well defined. The physical picture of our mapping includes a BKT-like transition as a signature of the deconfinement of the endpoints of the strings.
Future extensions of our work include the study of extended models with $SU(2)$ spins coupled directly to fluctuating Néel domain walls, enabling a better analysis of how $SU(2)$ symmetry is broken on different length scales.
Parallel efforts will be devoted to the search for hidden AFM structures in ongoing quantum simulation experiments.
Their ability to take quantum-projective measurements with
full simultaneous spin and charge resolution will allow to directly observe hidden AFM order, even when two-point spin correlations decay on short length scales comparable to the lattice spacing.

\section{Data availability}
The data that support the findings of this article are not publicly available. The data are available from the authors upon reasonable request.

\section{Acknowledgements}
We acknowledge fruitful discussions with Immanuel Bloch, Annabelle Bohrdt, Philipp Preiss and  Reja Wilke.
This research was funded by the European Research Council (ERC) under the European Union’s Horizon 2020 research and innovation program -- ERC Starting Grant SimUcQuam (Grant Agreement No. 948141) and QuantERA II (Grant Agreement No. 101017733), by the QuantERA grant DYNAMITE, and by the Deutsche Forschungsgemeinschaft (DFG, German Research Foundation) under project number 499183856 and under Germany's Excellence Strategy -- EXC-2111 -- project number 390814868.
This research is part of the Munich Quantum Valley, which is supported by the Bavarian State Government with funds from the High-tech Agenda Bayern Plus.

G.d.P. and G.D. contributed equally to this work. F.G. supervised the project. G.d.P., G.D. and S.L. performed the Monte Carlo simulations. All authors discussed the effective models and interpretation of results. 

\appendix 

\section{Motivation for the effective model} \label{sec:methods_VortexIntMotivation}

Here we construct our effective string model, Eq.~\ref{eq:Hamiltonian}. We start from the non-linear $\sigma$-model (\NLSM) description of the two-dimensional antiferromagnet~\cite{Chakravarty1989, Fradkin2013}. The Néel order parameter is represented by the vector field $\neelfield(\mathbf{x})$, with $|\neelfield| = 1$, and the energy functional reads
\begin{equation} \label{eq:nlsm}
    E_{\text{\NLSM}}[\neelfield] = \frac{\rho_s}{2} \int d^2x \, \left( \nabla \neelfield \right)^2,
\end{equation}
where $\rho_s$ denotes the spin stiffness.
As discussed in the main text, a string corresponds to a domain wall of microscopic width (comparable to the lattice spacing $a$) across which the Néel order shifts by $\pi$, i.e. $\neelfield \to -\neelfield$. For a given configuration of strings and vortices, we estimate the effective energy by minimizing $E_{\text{\NLSM}}$ over Néel field configurations compatible with these $\pi$-shift boundary conditions. 

Let us distinguish between energy costs that occur \textit{locally} within the discretization length scale $a$ around domain walls (i.e., the energy contributions arising from the stripes themselves), and those which occur outside of this range (i.e., the energy contributions from the field formed between the stripes). We begin with the local terms. Away from its endpoints, a domain wall yields an energy cost proportional to its length, 
due to the short-distance cost of the local $\pi$-shift associated with the string of holes bound to the wall. On the lattice, this produces the string tension term 
$- h \sum_{\langle \mathbf{i},\mathbf{j} \rangle } \tau^z_{\langle \mathbf{i},\mathbf{j} \rangle} $. 
Additionally, each vortex carries a fixed short-distance core energy arising from the region of size $\sim\!\!a$ around the end of the domain wall where the Néel field is locally varied to form a half-vortex. We parametrize this local contribution as $  - \mu \sum_{\mathbf{i}} n_{\mathbf{i}} $ and take $\mu < 0$, so that creating an endpoint costs positive energy.

What remains is to estimate the \textit{nonlocal} energy of the vortices. Consider the interaction between two string-endpoints separated by distance $d$, where $d$ is small compared to the correlation length  $\xi^*$ of the AFM background, and suppose that no other vortices are close to them or contribute to the interaction. 
In this regime, the pair of vortices is embedded in an ordered Néel background, and the energy-minimizing configuration can be chosen coplanar. Restricting the field to this plane, we write 
$\neelfield(\mathbf{x}) = \cos\theta(\mathbf{x}) \, \mathbf{e}_1 + \sin\theta(\mathbf{x}) \,\mathbf{e}_2$ 
with perpendicular directions $\mathbf{e}_1$, $\mathbf{e}_2$, where $\nabla \theta = \nabla \psi + \nabla \times \phi $ and $\nabla \times (\nabla \psi) = 0$. Eq.~\ref{eq:nlsm} reduces to $E_\text{\NLSM}[\phi] = \frac{\rho_s}{2} \int d^2x \, (\nabla \phi)^2$. A string is a branch cut across which $\phi$ changes by $\pi$, and around
each string-endpoint $\phi$ winds by $2\pi u_i = \pm \pi$.
The energy-minimizing field obeys 
$\nabla^2\phi = - \sum_i u_i \delta(\mathbf{x} - \mathbf{x}_i)$
in which the sum is taken over all domain walls ends $i$ with vorticities $u_i$ at positions $\mathbf{x}_i$. 
Away from the vortex cores and branch cuts, the energy-minimizing phase satisfies $\nabla^2\phi=0$, subject to the circulation conditions $\oint_{\mathcal{C}_i}\nabla\phi\cdot d\boldsymbol{\ell}=2\pi u_i$, where $u_i=\pm1$ and $\mathcal{C}_i$ is a closed path around a charge. A representative solution is $\phi(\mathbf{x})=-\sum_i u_i\arg(z-z_i)$, with $z=x+iy$, which is the phase field entering the correlator in Eq.~\ref{eq:U1_correlator}. 
Substituting into $E_\text{\NLSM}$, we obtain the logarithmic energy cost
proportional to $u_{\mathbf{i}}u_{\mathbf{j}} \, \mathrm{log}\vert \mathbf{i}-\mathbf{j} \vert $ (cf. integer-vortex calculations in the XY-model \cite{Polyakov1987, Fradkin2013}), up to local self-energies that are absorbed into $\mu$. Specifically, comparing with Eqn.~\ref{eq:Hamiltonian}, we identify the coefficient $M = 2\pi \rho_s$.
The interaction is attractive in the case of opposite vorticities, $u_{\mathbf{i}} = -u_{\mathbf{j}}$ at sites $i$, $j$, and repulsive for equal vorticities.

Altogether, we obtain the effective model in Eq.~\ref{eq:Hamiltonian}. This is applicable as long as the vortex density is sufficiently low, so that no other free string-endpoint lies at a comparable distance from the considered pair. At larger separations or in a dense configuration, the logarithmic coupling should not be interpreted as an exact interaction; instead, other vortices effectively screen the long-distance field. To keep the model minimal, we nevertheless include the logarithmic coupling between all vortex pairs as an approximation. Note further that, strictly speaking, Eq.~\ref{eq:Hamiltonian} should be viewed as an effective classical free-energy functional for the string degrees of freedom, even though we write it as a Hamiltonian. The configurational entropy of vortex positions and string arrangements is not added as a separate term, but it is generated automatically by the statistical ensemble over $\tau^z$ and $u_i$. Finally, note that in a finite system the coupling should be normalized as $\mathrm{log}(\vert \mathbf{i}-\mathbf{j} \vert / L) $ rather than $\mathrm{log} \vert \mathbf{i}-\mathbf{j} \vert $. At a fixed system size, this can be absorbed into a shift of the vortex potential $\mu$. When comparing different system sizes with varying vortex numbers, the same normalization must be used consistently, as the additive $-\log L$ contribution alters the effective vortex energy. In our simulations, we only vary $L$ for the finite size analyses where the vortex number is fixed, so this does not affect the sampled relative weights.

\section{$U(1)$ correlator for BKT phase transitions} \label{sec:methods_U1op}
For tracking the BKT-type deconfinement transition, we define an order parameter which measures the correlation of phases acquired from the open ends of domain walls in the \NLSM, defined as in Eq.~\eqref{eq:U1_correlator}.

The angular field that we consider, $\phi(\mathbf{x})$, is generated by the vortices through the function $\Im\{\log(\mathbf{x}-\mathbf{x}_j)\}$, which is the harmonic conjugate of the two-dimensional Laplacian Green function~\cite{arfken_mathematical_2013, brown_complex_2009}.

In two dimensions, the fundamental solution of the unit-point source Laplace equation is given by $\nabla^2G(\mathbf{x} - \mathbf{x}') = \delta^{(2)}(\mathbf{x} - \mathbf{x}')$ where $G(\mathbf{r}) = \frac{1}{2}\pi \log(\frac{|\mathbf{r}|}{a})$. 
Accordingly, for a set of point charges \(u_j\) located at \(z_j\), the scalar potential generated by these sources is obtained by superposition as $ \Phi(z) \propto \sum_j u_j \log|z-z_j|$.
In our case, however, the relevant phase field is not constructed directly from this scalar Green-function kernel, but rather from its harmonic conjugate, $\phi(z) = \sum_j (-u_j)\,\arg(z-z_j)$ (the overall sign depends on the chosen convention for the vortices).

We compute this correlation function directly from snapshots produced by the Monte Carlo simulations. At each site of the lattice, we evaluate the angular contribution of every string-endpoint and then bin the resulting correlations according to the Manhattan distance between the two sites.
Because finite-size effects are more prominent near the boundary, we only consider the correlation between sites near the center of the system (i.e., with at least distance $L/4$ from the boundary). Results of this analysis are presented in Fig.~\ref{fig:Monte_Carlo_results}d, where we show the clear power law to exponential law transition for exemplary parameters. The error bars are calculated with the jackknife procedure and are considered for the fit of the curves. In Sec.~\ref{sec:Benchmark_U1} of the Appendix, we test the validity of the correlator in an $XY$-model with direct BKT phase transition.

\section{Modeling nematic order} \label{sec:methods_stripes}
\begin{figure}[t!]
    \centering
    \includegraphics[width=1.0\linewidth]{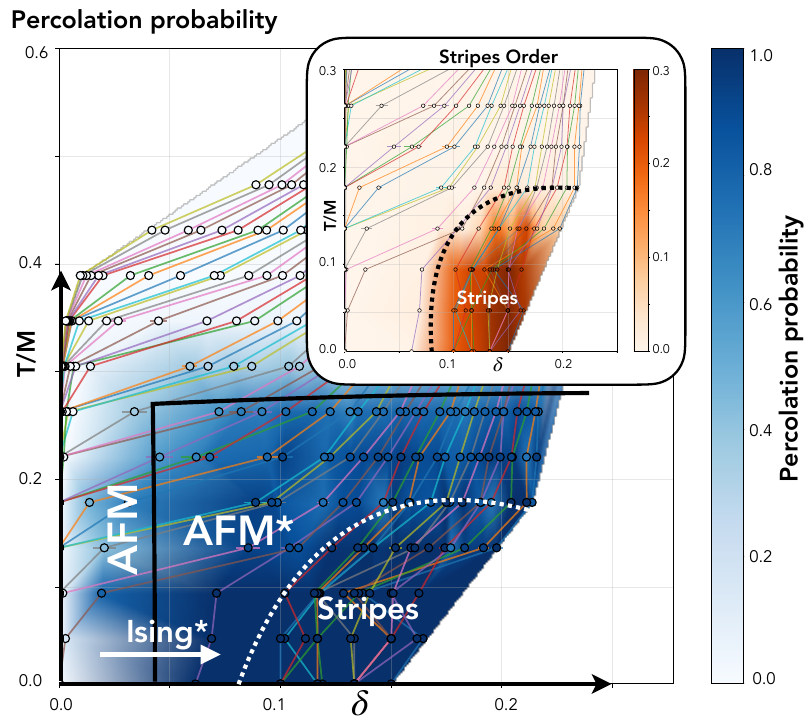}
    \caption{\textbf{Stripe order simulations.} Plot of the percolation probability for different points in the temperature vs. density phase diagram obtained from Monte Carlo simulations of the system with Hamiltonian~\eqref{eq:StripeHamiltonian} with system size $L=30$. The connected lines correspond to Monte Carlo sweeps at different fixed $h$. The other parameters are fixed: $V_{\mu}=0.15$ and $W_{\mu}=1.5$, $M=1.0$ and $J_2=3.0$.
    The continuous color background is extrapolated from the discrete points, while the hole density is calculated from the string density. The stripe order parameter defined in Eq~\eqref{eq:stripe_op} is plotted in the temperature-vs-density phase diagram in the inset and shows an onset of the symmetry broken phase inside the percolating region.}
    \label{fig:Stripes_Order}
\end{figure}
In Sec.~\ref{sec:eff_string_model}, we noted that the minimal model studied here can be generalized by introducing additional effective interaction terms. This provides a flexible ansatz to investigate a broader class of effective theories. As an illustrative example, we consider a modified model which, among other features, exhibits (spontaneous) $C_4$ symmetry breaking and incorporates further-range interactions between fluctuating stripes. This setup is particularly interesting as a possible framework for modelling stripe formation together with spontaneous \(C_4\)-symmetry breaking, phenomena that would be of considerable interest for potential experimental implementations.

For notational convenience and to make the energetic contributions clearer, we define
\begin{equation}
\sigma^z_{\langle \mathbf{i},\mathbf{j} \rangle}=\frac{1-\tau^z_{\langle \mathbf{i},\mathbf{j} \rangle}}{2},
\end{equation}
with $\sigma^z \in \{0,1\}$. This reformulation simply corresponds to a reparameterization of the Hamiltonian. The resulting model reads
\begin{equation} \label{eq:StripeHamiltonian}
\begin{split}
     \mathcal{H} = & \sum_{\alpha=x,y} h_{\alpha} \sum_{\langle \mathbf{i},\mathbf{j} \rangle_\alpha}
    \sigma^z_{\langle \mathbf{i},\mathbf{j} \rangle} - \mu \sum_\mathbf{i} n_\mathbf{i}
     \\
    & - \sum_{\alpha=x,y} V_{\alpha}
    \sum_{\langle \mathbf{i},\mathbf{j},\mathbf{k} \rangle_\alpha}
    \sigma^z_{\langle \mathbf{i},\mathbf{j} \rangle}
    \sigma^z_{\langle \mathbf{j},\mathbf{k} \rangle}
    + J_2 \sum_{\mathbf{j}}
    \prod_{\langle \mathbf{i},\mathbf{j} \rangle}
    \sigma^z_{\langle \mathbf{i},\mathbf{j} \rangle} \\
    & + \sum_{\alpha=x,y} \sum_{\mathbf{j} \neq \mathbf{i}}
    \frac{1}{2} W_{\alpha}(\vert \mathbf{j}-\mathbf{i} \vert)
    \sigma^z_{\langle \mathbf{j},\mathbf{j}+\mathbf{e}_{\alpha} \rangle}
    \sigma^z_{\langle \mathbf{i},\mathbf{i}+\mathbf{e}_{\alpha} \rangle}\\
    & - \frac{1}{2}M \sum_{\mathbf{j} \neq \mathbf{i}}
    n_{\mathbf{i}}n_{\mathbf{j}}
    u_{\mathbf{i}}u_{\mathbf{j}}
    \log\bigl(\vert \mathbf{i}-\mathbf{j} \vert \bigr).
\end{split}
\end{equation}
Here, $V_\alpha$ denotes the local alignment interaction, while $W_\alpha$ controls the long-range interaction between vertical and horizontal strings and we generally choose $h_{\alpha} = h$, $V_{\alpha} = V$ and $W_{\alpha} = W$. ${\langle \mathbf{i},\mathbf{j} \rangle}_{\alpha}$ represents two neighboring vertices ($\mathbf{i}$ and $\mathbf{j}$) aligned in the direction $\mathbf{e}_{\alpha}$ and $\langle \mathbf{i},\mathbf{j},\mathbf{k} \rangle_\alpha$ indicates three vertices aligned in the direction $\mathbf{e}_{\alpha}$, with $\mathbf{k}$ the common vertex of the two links.

This Hamiltonian can be tuned to simulate mixed-dimensional systems, for example, by allowing long-range interactions only between vertical strings, or to investigate nematic responses. In such a case, the interaction may be chosen as
\begin{align}
  W(\mathbf{r}) =
  \begin{cases}
    W_0 e^{-r^2/\sigma_W^2} & \quad \text{if } \alpha = y,\ \mathbf{r} = r\mathbf{e}_x, \\
    0 & \quad \text{otherwise}.
  \end{cases}
\end{align}
Here, $W_0$ is a constant prefactor, while the interaction profile is modulated by a Gaussian of width $\sigma_W$.

To make the Hamiltonian explicitly $C_4$ symmetric, the further-range interaction must be present with equal strength for both vertical and horizontal strings. We therefore choose
\begin{align}
  W(\mathbf{r}) =
  \begin{cases}
    W_y e^{-r^2/\sigma_W^2} & \quad \text{if } \alpha = y,\ \mathbf{r} = r\mathbf{e}_x, \\
    W_x e^{-r^2/\sigma_W^2} & \quad \text{if } \alpha = x,\ \mathbf{r} = r\mathbf{e}_y.
  \end{cases}
\end{align}
By varying these interaction strengths, one can probe nematic responses and possible symmetry-breaking instabilities.
In addition, the term
\begin{align}
  J_2 \sum_{\mathbf{j}}
  \prod_{\langle \mathbf{i},\mathbf{j} \rangle}
  \frac{1-\tau^z_{\langle \mathbf{i},\mathbf{j} \rangle}}{2},
\end{align}
penalizes configurations in which vertical and horizontal stripes cross. Equivalently, this term suppresses the formation of stars in the stripe configurations. This additional constraint may be important for future Monte Carlo studies of quantum fluctuations in this model.

In Fig.~\ref{fig:Stripes_Order}, we show the percolation probability for a representative set of parameters: $V_{\mu}=0.15$ and $W_{\mu}=1.5$, $M=1.0$ and $J_2=3.0$. The precise choice of parameters is not essential, since several parameter sets display the same qualitative behavior. The extended Hamiltonian exhibits percolation and confinement regimes that are qualitatively similar to those observed in the minimal model without stripe-alignment terms, indicating that these features are robust against the inclusion of stripe interactions. In addition, we observe a nematic-order \textit{dome} at low temperatures, which we detect using the following nematic orientation order parameter:
\begin{align}\label{eq:stripe_op}
  \Bigg| \frac{n_{v}^s}{n_{v}^s + n_{v}^{ns}} - \frac{n_{h}^s}{n_{h}^s + n_{h}^{ns}} \Bigg|,
\end{align}
where we define $n_{v}^s$ $(n_{h}^s)$ the total number of vertical (horizontal) links with $\tau^z = -1$ and $n_{v}^{ns}$ $(n_{h}^{ns})$ the total number of vertical (horizontal) links with $\tau^z = 1$.
These results suggest that the generalized model provides a promising framework for future studies of stripe formation and discrete symmetry breaking.

\section{Monte Carlo simulations} \label{supp_sec: numerical details}
\begin{figure*}[t!]
    \centering
    \includegraphics[width=.8\linewidth]{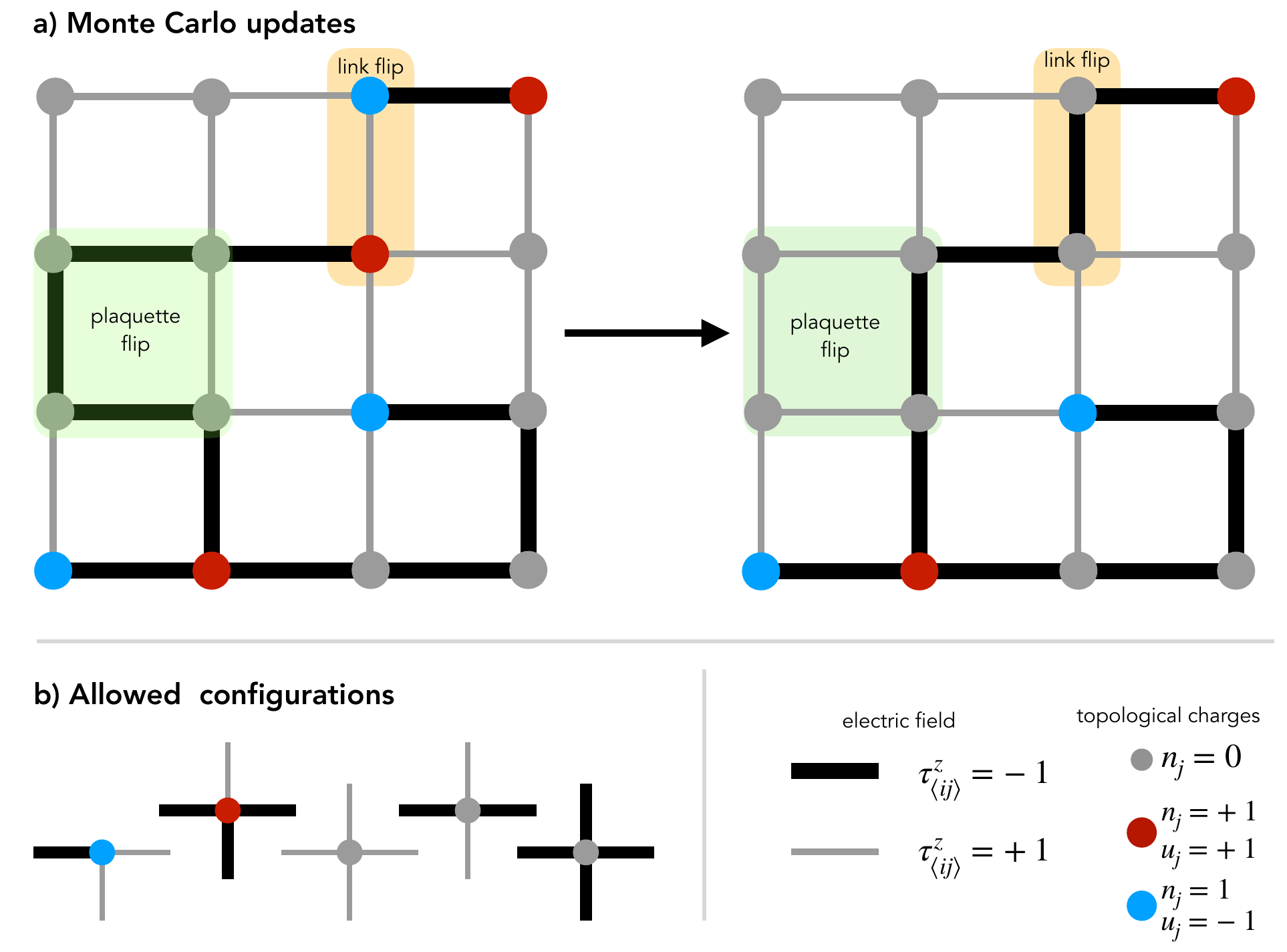}
    \caption{\textbf{Update and structure of the Monte Carlo analysis.} \textbf{a)} We show the lattice configuration before~(left) and after~(right) the update procedures for the Monte Carlo sampling are shown. In orange, a link-flip update either creates or annihilates a couple of charges and flips the link in between or moves a charge to an empty  site, flipping the string accordingly. The plaquette update, in green, flips all the links of a square plaquette, leaving the charge configuration untouched. Through the simulation, the total vortices is conserved, while the total number of charges can change. \textbf{b)} The Gauss's law constraints are shown.}
    \label{fig: MC_updates}
\end{figure*}
We employ classical Markov chain Monte Carlo (MCMC) simulations to sample the equilibrium probability distribution associated with the model defined in Eq.~\eqref{eq:Hamiltonian}. From these simulations, we then obtain the expectation values of the energy and of the percolation order parameter $\Pi$, and characterize their statistical behavior as functions of the control parameters $h$, $\mu$, and $M$. The sampled configurations are subsequently used to evaluate the $U(1)$ correlations via Eq.~\eqref{eq:U1_correlator}.
The Monte Carlo code is implemented in C++ using the Boost C++ libraries. The lattice is represented as a graph and data processing and post-processing are done in Python using the NumPy \cite{Harris2020}, SciPy \cite{Virtanen2020}, multiprocessing, and Matplotlib \cite{Hunter2007} libraries.

The configuration space is composed of classical degrees of freedom $\tau$ defined on a two-dimensional square lattice of linear size $L$, with a total number of sites $N = L^{2}$, and subject mixed boundary conditions. The variable $\tau$ is binary and represents the occupation state of a lattice edge: $\tau = 1$ denotes an empty edge, whereas $\tau = -1$ denotes an edge occupied by a flipped link. In addition, each lattice node carries a vortex number $n_i = 0,1$ and a vorticity degree of freedom, which can take values $\pm 1/2$, or $0$ when no vortex is present.
The sampling procedure is based on a Markov chain designed to satisfy detailed balance and ergodicity. Candidate configurations are generated through a set of local update moves and are accepted or rejected according to the standard Metropolis--Hastings criterion. These updates include local link flips which additionally create or annihilate pairs of vortices at the endpoints of the selected link. In this way, the total topological charge is conserved globally, while vortices are allowed to propagate throughout the system. Plaquette updates are also included, where all the links belonging to a chosen plaquette are flipped.
\begin{figure*}[t!]
    \centering
    \includegraphics[width=1.\linewidth]{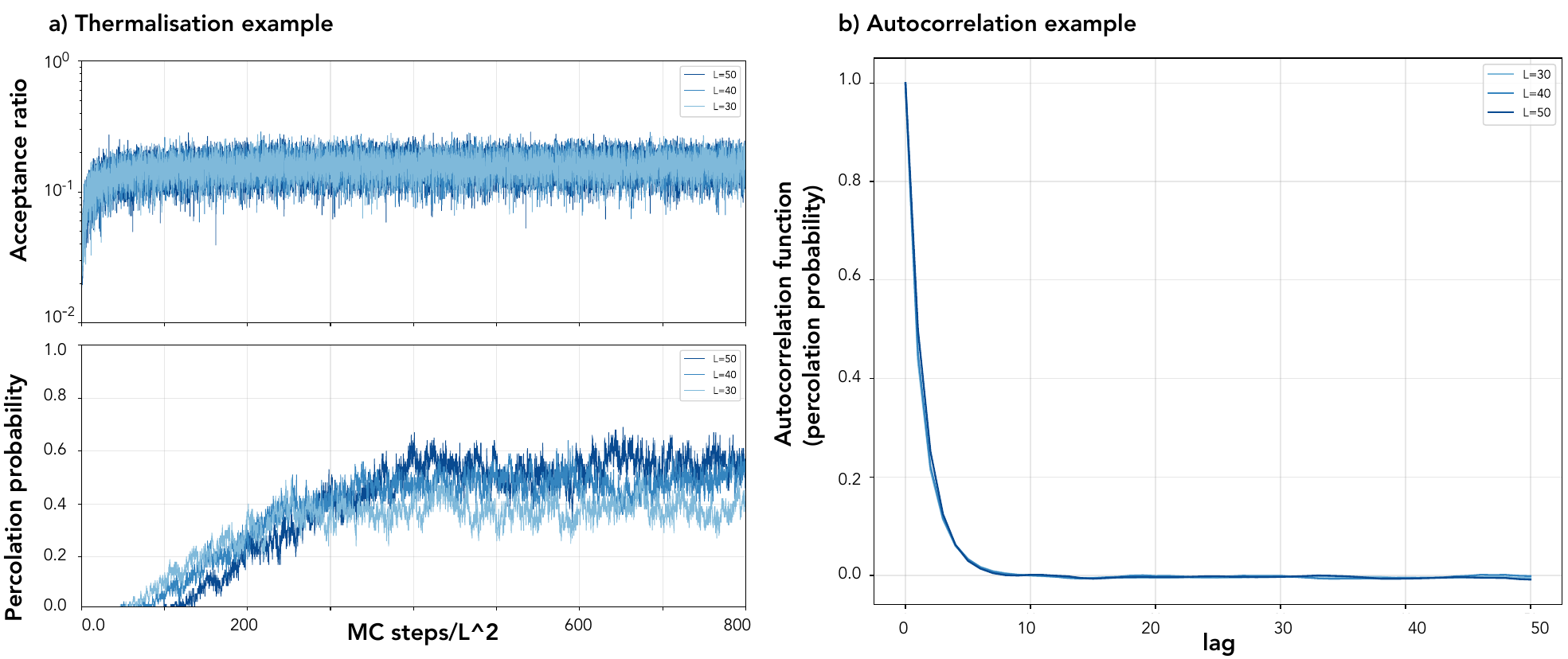}
    \caption{\textbf{Thermalization and autocorrelation.} We show some numerical details of the Monte Carlo runs. \textbf{a)} We plot the thermalization curves of the percolation probability and the acceptance ratio. The data are averaged over 10 runs at $T/M = 0.15$ and $h/M = 0.06$, i.e., inside the percolating phase, and we confirm that the system thermalizes after $400×L^2$ steps. \textbf{b)} We show the integrated autocorrelation time (we call it simply "autocorrelation") of the percolation probability for the same regime of parameters. We again plot the average over 10 runs and find a small autocorrelation.} 
    \label{fig: MC_autocorrelations_thermalisation}
\end{figure*}

Recently, POPs have been introduced in the study of electric confinement in $\mathbb{Z}_2$ lattice gauge theories~\cite{Linsel2024}. These operators quantify the winding number associated with connected clusters $C$ which are formed by neighboring links $l$ for which $\tau^z=-1$ and that have vertices in common, i.e., $\forall l \in C \,\ \tau^z_l=-1$. Specifically, the percolation probability is defined as the expectation value of the projector
\begin{equation}
    \Pi = \sum_{W(j)\neq 0} \left| \{ \tau^z \}_j \right\rangle \left\langle \{ \tau^z \}_j \right|,
\end{equation}
where $W(j)$ is the non-zero winding number, meaning for each analyzed snapshot it is the percolation indicator, a binary quantity $W(j)= 0 \lor 1$ is calculated and then averaged to get the final value of $\Pi$.
For the evaluation of the POPs, we simulate periodic systems up to $L^2=50^2$ in terms of unit cells at temperatures varying in the range from $0.01$ to $1.5$ $M$ and take up to $10^4$ snapshots for every data point. We repeat our parameter scans, measure the POPs and $U(1)$ order parameter, and other observables, such as the energy, for every snapshot.
Statistical uncertainties from the MCMC are estimated using integrated autocorrelation times or bootstrap techniques~\cite{Politis2004, Politis1994, Linsel2024}, with autocorrelation times explicitly taken into account.

The simulations are based on the Hamiltonian given in Eq.~\eqref{eq:Hamiltonian}. The energy $H(\tau^z_{\langle \mathbf{i},\mathbf{j} \rangle}, u_{\mathbf{i}})$ is completely specified by a given configuration of the strings $\tau^z_{\langle \mathbf{i},\mathbf{j} \rangle}$ and the topological charges $u_{\mathbf{i}}$. Sampling is carried out with the Metropolis--Hastings algorithm, combining move updates and plaquette updates, as illustrated in Fig.~\ref{fig: MC_updates}a.
Because we work in the grand-canonical ensemble, the number of strings in the system as well as the number of vortices must be controlled through fixed chemical potentials: $h$ acts as such for the strings and $\mu$ for the endpoints. The achievable resolution in the density of vortices is determined by the system size, since this controls how finely $\mu$ can be tuned.
We restrict the theory to the configurations shown in Fig.~\ref{fig: MC_updates}b, which corresponds to considering an open end as the extremity of a cluster where an odd number of strings meet at a vertex. In this setting, $n_j$ can be expressed as in Eq.~\eqref{eq:top_charges}.
The Monte Carlo update procedure is particularly straightforward in the grand-canonical formulation that we use. This setup allows the creation and annihilation of charges without generating unphysical configurations, provided that the total topological charge remains zero. The first type of update, which may change the total number of vortices, is the link flip, which is proposed by randomly selecting a link and flipping the value of $\tau_{\langle \mathbf{i}, \mathbf{j}\rangle}$. To make this update gauge invariant, we also randomly select the topological charges on the two vertices the link connects, $\mathbf{i}$ and $\mathbf{j}$, such that the total charge remains invariant after the update, so $u_i + u_j = 0$ (i.e. $u_i=+1/2$ and $u_j=-1/2$ or $u_i=+1/2$ and $u_j=-1/2$ with $50\%$ probability) and the update is accepted only if the neutrality is conserved $\sum_i u_i = 0$. 
This mechanism allows for updates that effectively are either a pair creation or annihilation, with associated gauge-invariant link flip, or behave as a step in one direction of one of these vortices, with associated modification of the connected string.
The plaquette update, by contrast, leaves the number of vortices unchanged and can therefore always be applied in this framework. Since neither update involves an asymmetric trial probability distribution, standard Metropolis sampling can be used. At each Monte Carlo step, a plaquette update or an electric-field flip update is proposed with probability $50\%$ each. 

Each run begins with a thermalization stage of $400 \times L^2$ Monte Carlo steps. After thermalization, $10^4$ samples are recorded, with consecutive samples separated by $2 \times L^2$ Monte Carlo steps. Representative data for the thermalization and autocorrelation behavior at $T/M =0.15$ and $ h/M = 0.06$ are shown in Fig.~\ref{fig: MC_autocorrelations_thermalisation}b for the three system sizes used, namely $30$, $40$, and $50$. Autocorrelations between sampled configurations are taken into account in all quoted error bars for directly measured observables, such as percolation probabilities and Binder cumulants.
The snapshots are generated on a periodic link-lattice, meaning that the interactions between the edges are taken to be periodic, but a nonperiodic potential is used for the vortex interactions.
We use PBC for percolation analysis, as it provides better characterization of the percolating regimes and yields improved results (the clusters are simply longer).
Under PBC, the percolation probability and percolation strength are determined by checking whether a string configuration forms a path that winds around the system in at least one spatial direction. The code we use performs a depth-first search starting from a selected set of boundary sites and continues the exploration only in string-occupied links. We assign a winding number to every vertex that has already been visited we update it whenever the path crosses a predefined cut through the lattice. Once already visited vertices are encountered again and the winding numbers are different, the percolation in such direction is detected since it indicates the existence of a non-contractible winding path.
This definition is preferable to the standard open-boundary criterion because it exhibits smaller finite-size effects. In particular, the open-boundary definition is subject to corrections of order $\mathcal{O}(1/L)$, while the winding-based criterion used here reduces them to $\mathcal{O}(1/L^2)$~\cite{Ziff1992, Newman2000}. This yields a significant improvement in the finite-size scaling analysis.

The Monte Carlo updates responsible for generating or moving vortices are implemented with open boundary conditions, and distances are measured using the usual Euclidean, non-periodic metric. This choice follows from the definition of the order parameter in Eq.~\eqref{eq:U1_correlator}, which is naturally formulated with open boundaries. The formulation of a PBC-adapted order parameter would allow the complete homogenization of the boundary conditions.

\section{Compressibility} \label{supp: compressibility}
\begin{figure*}[t]
    \centering
    \includegraphics[width=1.0\linewidth]{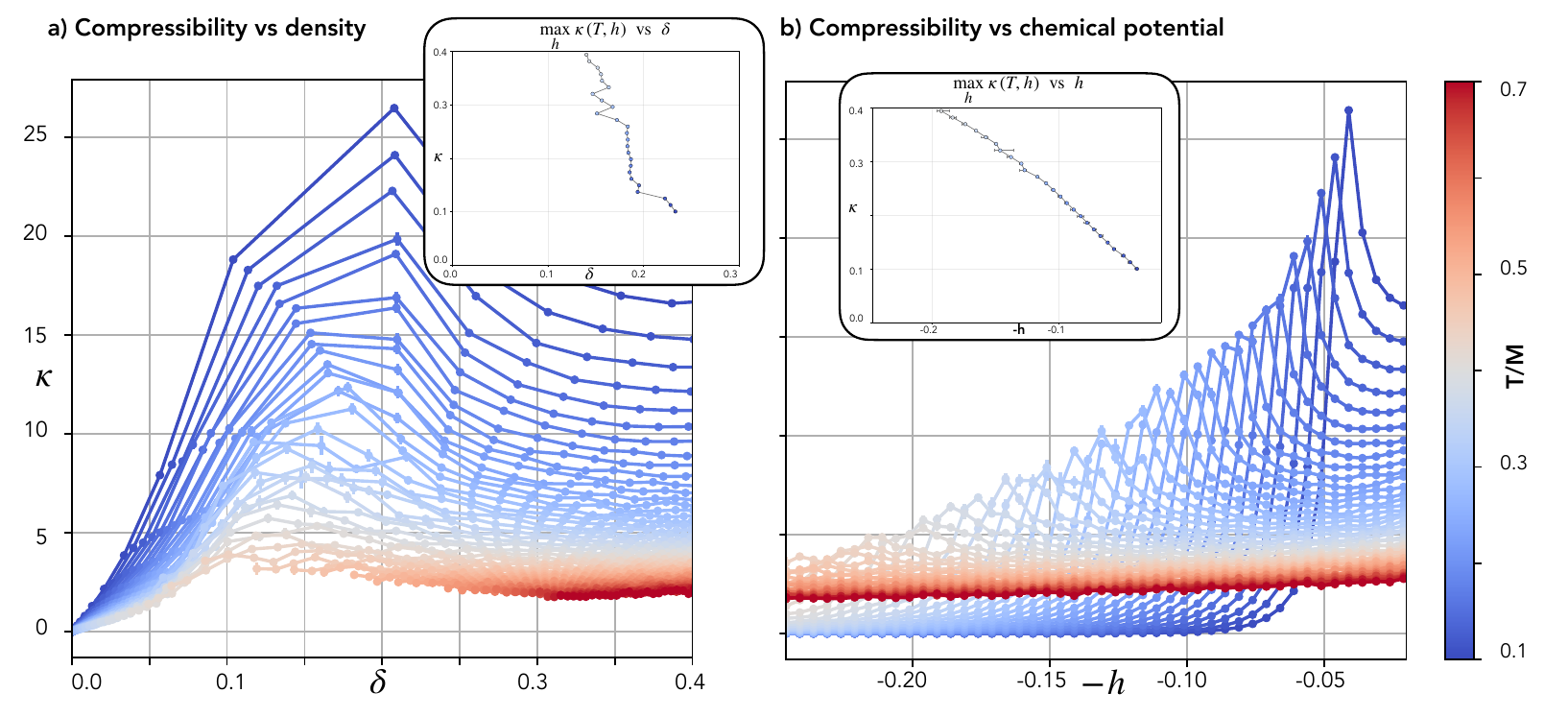}
    \caption{\textbf{Compressibility plots.} \textbf{a)} Compressibility curves as a function of the inferred hole density, in the inset we show the Widom line extracted from the maxima of \(k(h,T)\) at fixed temperature represented in the density-temperature plane. We do gaussian interpolation on the curves to extract the maxima. \textbf{b)} Compressibility curves in function of the chemical potential $h$, in the inset we show the Widom line. Each point marks the chemical potential where the compressibility is largest for that temperature. The choice to plot the values with $-h$ on the x-axis is only done to match the definition of Ref.~\cite{Sordi2012}.}
    \label{fig: detailed_compressibility}
\end{figure*}
Many prior works have considered the charge compressibility and the Widom line, since these provide particularly useful diagnostics for the pseudogap crossover, connecting interesting spectroscopic behaviors to underlying thermodynamic changes of structure~\cite{Kotliar2002, Sordi2010, Sordi2011, Sordi2012}. In the interpretation of Sordi \emph{et al.}~\cite{Sordi2012}, the pseudogap temperature $T^*$ is not simply identified from the suppression of low-energy spectral weight, but is instead associated with the continuation of a first-order transition between a pseudogap metal and a correlated Fermi liquid. In this context, the charge compressibility
\begin{equation}\label{supp-eq:compressibility}
    \kappa = \frac{1}{n^2}\left(\frac{\partial n}{\partial h}\right)_T    
\end{equation}
carries interesting additional information, with pronounced maxima near the critical regimes, which can be used to draw the Widom line and progressively sharpen upon approaching criticality. 
In Ref.~\cite{Sordi2012}, the Widom line is therefore the line where the maxima of different response functions converge.
Importantly, this has led to interpreting $T^*$ as a thermodynamically meaningful point: the same line that organizes anomalies in $\kappa$ also coincides with rapid changes in the density of states, spin susceptibility, and singlet correlations. 
In that context, the Widom line is hence a natural marker of the pseudogap crossover, as it identifies the location of a collective signal seen in multiple observables of the influence of the hidden Mott-driven critical point~\cite{Sordi2012}.

Because our model is purely effective in nature, we do not have direct access to more microscopic degrees of freedom such as spins and holes. Instead, the only explicit variables retained in the description are the link variables, while all remaining microscopic information has been integrated out in a consistent manner. As a consequence, in our system, the Widom line cannot be identified through observables associated with those microscopic sectors, but must instead be inferred from the behavior of the compressibility. In practice, we locate it by determining the maxima of the rescaled compressibility function that we obtain from our effective model. To obtain the appropriate counterpart of the quantity defined in Eq.~\eqref{supp-eq:compressibility}, we use the effective hole density computed from Eq.~\eqref{eq:density} together with the chemical potential of the string variables, $h$, appearing in Eq.~\eqref{eq:Hamiltonian}.

To obtain the formula for the compressibility in our effective description, we start by considering the specific involved energy term:
\begin{equation}
    \mathcal{H} = -h \sum_{\langle i, j \rangle} \tau_{\langle i, j \rangle}.
\end{equation}
As explained, $\tau_{\langle i, j \rangle} = -1$ indicates that the single link must be considered a string. We explicitly write the hole-doping dependence as $\langle \hat\tau_{l} \rangle = 1 - \langle 2\delta_l \rangle $, and 
\begin{equation}
    \langle  \mathcal{H} \rangle = -\frac{h}{N_l} \sum_{\langle i, j \rangle}  1 -2\delta_{\langle i, j \rangle}  
\end{equation}
which, on average, contributes as 
\begin{equation}
    -h \frac{N_l}{N_l} + 2h\sum_{\langle i, j \rangle} \frac{ \delta_{\langle i, j \rangle} }{N_l}.
\end{equation}
We consider $n_e= 1-\delta$, where $n_e$ is the electron-density. For example, in the Mott insulator, we would have $n=1$ and $\delta=0$.
Therefore, the formula used in the calculations of the compressibility is:
\begin{equation}
    \frac{1}{n_e^2}\left(\frac{\partial n_e}{\partial h}\right)_T  = -\frac{1}{(1-\delta)^2}\left(\frac{\partial \delta}{\partial h}\right)_T .
\end{equation}
In our simulation, the derivative is obtained through the ratio of discrete differences of values of $n$ and $h$ at a fixed temperature slice. The Widom line tracks the points $h_{\max}(T)=\max_h \kappa(T,h)$, which are extracted at fixed temperature.

In Fig.~\ref{fig: detailed_compressibility}, we show an example of a compressibility check where the parameter sweep is taken densely in both temperature and $h$ regimes. We set $L=30$, $M=1$ and $T$ in the range $[0.1,1]$, $h$ in the range $[0.01, 0.25]$. Fig.~\ref{fig: detailed_compressibility}a shows the compressibility $\kappa$ plotted against the density of holes, and the inset shows the Widom line as defined. Fig.~\ref{fig: detailed_compressibility}b shows the same quantity against the chemical potential~$h$. 
The lines are not smooth because of the limited amount of simulations used.
Unlike the Widom-line construction of Ref.~\cite{Sordi2012}, where extrema of several response functions converge, here we use only the results of the compressibility.

\section{Finite-size analysis for the percolation transition}\label{supp_sec: finite-size analysis}
To characterize the percolation transition, we perform a finite-size analysis of the percolation probability. 
In particular, we examine selected sets of fixed-\(h\) simulation data shown in Fig.~\ref{fig:Monte_Carlo_results}, focusing on the evolution of the percolation curves with increasing system size. To identify the transition, we also analyze the Binder cumulant associated with the percolation strength, $P$, and determine its crossing point for different system sizes. The percolation strength is a quantity that generalizes the information provided from the percolation probability, since it also depends on the dimension of the string-clusters that percolate. Specifically, we consider:
\begin{equation}
U_P(h,L) = \frac{\langle P^4(h,L)\rangle}{\langle P^2(h,L)\rangle^2}
\end{equation}
whose crossing provides an estimate of the critical hole density \(\delta_c\) at which the percolation transition occurs. An example of this analysis is shown in Fig.~\ref{fig:Monte_Carlo_results}b, where we fix \(h=0.05\), vary \(T\) in the interval \([0.08,0.5]\), and consider the system sizes \(L=30,40,50\).

\section{Benchmark for the $\mathbf{U(1)}$ order parameter}\label{sec:Benchmark_U1}
\begin{figure*}[t!]
    \centering
    \includegraphics[width=1.\linewidth]{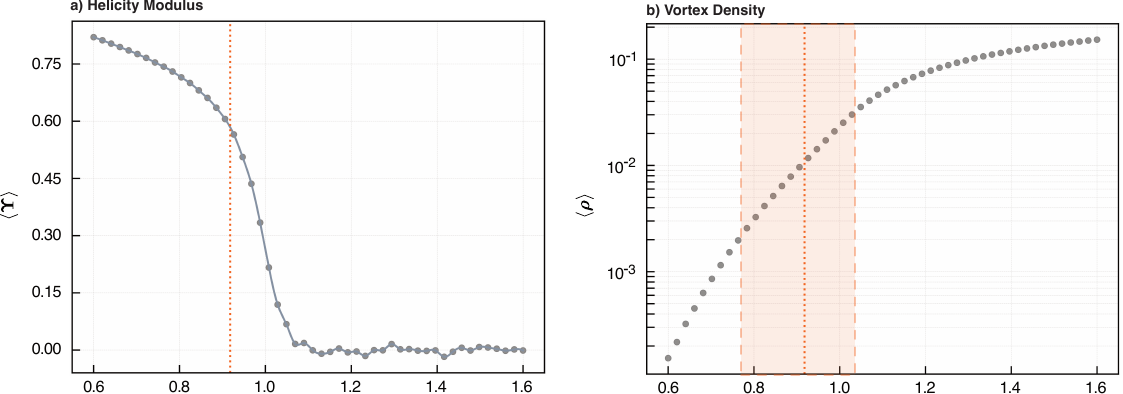}
    \caption{\textbf{Monte Carlo simulations of the XY model. a)} Helicity modulus $\langle \Upsilon \rangle_{x}$ for a system of size $L = 60$ with periodic boundary conditions and a perturbing field along x. The dashed orange line marks the critical temperature estimated from the Nelson-Kosterlitz criterion. 
    \textbf{b)} Vortex density $\langle \rho \rangle$ computed from the same simulation as in (a), with the dashed orange line marking the same critical temperature. 
    The rapid increase, in particular the exponential growth near the transition (orange region), corroborates the vortex proliferation characteristic of a BKT transition and the reliability of the vortex-detection algorithm.}
    \label{supp: Helper-Figure}
\end{figure*}
\begin{figure*}[h!]
    \centering
    \includegraphics[width=1.\linewidth]{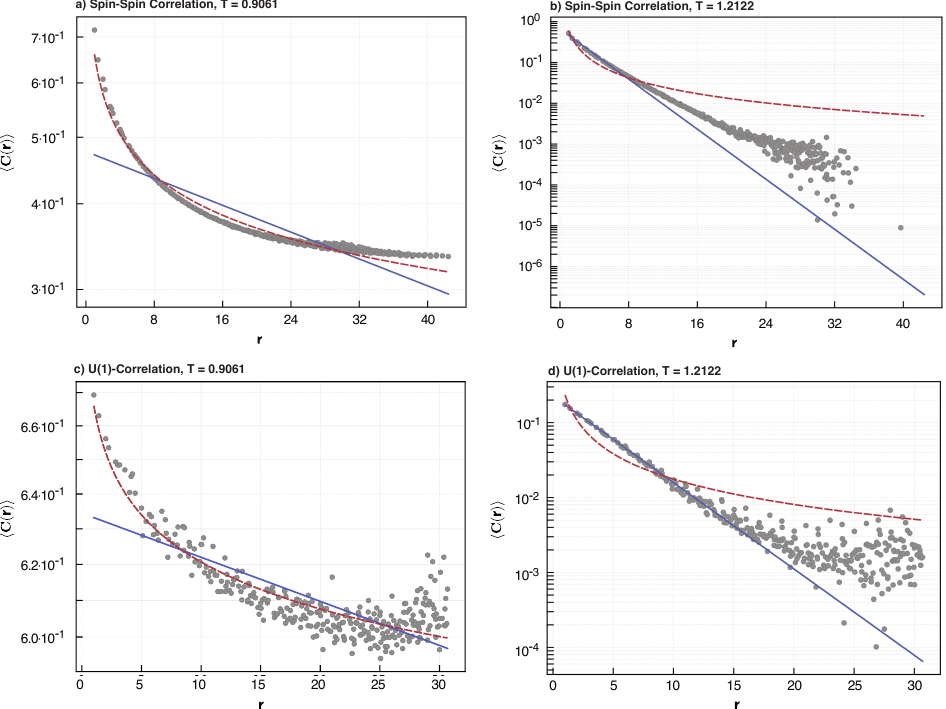}
    \caption{\textbf{Benchmark of the $\mathbf{U(1)}$ correlator in the XY model. a)} Spin-spin correlation function on $L = 60$ with periodic boundary conditions. The algebraic decay of correlations indicates the quasi-long-range-ordered phase. \textbf{b)} Spin-spin correlation function in the disordered phase, characterized by an exponential decay. The deviations from exponential decay at large distances arise from finite-size effects and should be accounted for when numerically precise results are required. Correlation distances are limited to $\sqrt{2} \cdot L/2$ due to periodic boundary conditions. \textbf{c)} $U(1)$ correlation function in the quasi-long-range-ordered phase, exhibiting algebraic decay. \textbf{d)} $U(1)$ correlation function in the disordered phase, showing exponential decay, with deviations at large distances arising from finite-size effects, as in panel (b). The $U(1)$ correlations are computed from the same Monte Carlo data as used for panels (a) and (b), using only vortices within the center $30 \times 30$ bulk region with open boundary conditions. Here, we impose a cutoff at $3/4$ of the maximum distance, again $\sqrt{2} \cdot 30$ due to the open boundary conditions, since the tail is dominated by numerical noise.}
    \label{supp: Correlation-Figure}
\end{figure*}
We benchmark the $U(1)$ correlator of Eq. (\ref{eq:U1_correlator}) in the two-dimensional classical XY model, where the BKT transition can be identified independently from the spin-spin correlation function and the helicity modulus.
To this end, we conduct classical Monte Carlo simulations of the Hamiltonian 
\begin{equation}\label{eq:xy-model}
    \mathcal{H}_{\text{XY}} = -J \sum_{\langle i, j \rangle} \textbf{S}_{i} \textbf{S}_{j} = -J \sum_{\langle i, j \rangle} \cos \left( \theta_{i} - \theta_{j} \right) \ ,
\end{equation} 
with $\textbf{S}_{i} = \left(S^{x}_{i}, S^{y}_{i} \right)^{T}$, 
$|\textbf{S}_{i}| = 1$, and $\theta \in [0, 2\pi)$.
Near the transition temperature, we employ the Wolff cluster update algorithm~\cite{Wolff-Algorithm_1989}. The remaining temperature steps are carried out using the Metropolis update algorithm~\cite{Metropolis1953}. 
We fix the coupling strength to $J=1$. The simulations are performed on a lattice with size $L = 60$ over the temperature range $T \in [0.6, 1.6]$ using $50$ equidistant steps. Each simulation consists of $5 \cdot 10^3$ thermalization steps followed by $10^6$ additional sweeps, from which $2 \cdot 10^4$ snapshots were recorded for analysis. 
Periodic boundary conditions (PBC) were applied throughout. However, for the computation of the $U(1)$ correlation, only vortices located within the $L/2 \times L/2$ bulk were taken into account, allowing for the use of open boundary conditions (OBC).

The computation of the $U(1)$ phase $\phi(\textbf{x})$, see Eq.~\eqref{eq:U1_correlator}, requires the coordinates of the elementary vortices, i.e., vortices with a winding number $n = \pm 1$. 
A plaquette is a square-shaped closed path on the real lattice, along the vertices between sites. The center of a plaquette defines a site of the dual lattice. We consider elementary plaquettes, denoted by $\Box$, corresponding to the smallest closed paths on the lattice and consisting of four spins.
An elementary vortex at dual-lattice site $k$ is identified by its winding number, which is obtained from the discrete closed path integral
\begin{equation}\label{eq:winding-number}
    n_k = \frac{1}{2\pi} \sum_{(i, j) \in \Box^{\circlearrowleft}_{k}} \Delta \theta_{ij} \ \ \ \ \text{with} \ \ \ \ \Delta \theta_{ij} := \theta_j - \theta_i  \ , 
\end{equation}
where the sum is taken counterclockwise along the plaquette, denoted by $\Box^{\circlearrowleft}_{k}$, and $\theta_j$ is the successor of $\theta_i$ along this path, see also Refs.~\cite{Berezinskii_1971, Kosterlitz1973}. To ensure that the correct local angle difference is obtained, we wrap $\Delta \theta_{ij}$ to the interval $[-\pi, \pi)$.
The vortex density $\langle \rho \rangle $ serves as a diagnostic quantity for assessing the reliability of the vortex-detection procedure, since the number of vortices is expected to increase rapidly with temperature, particularly in the vicinity of the critical point~\cite{Andrade25}.
We define the vortex density
\begin{equation}
    \langle \rho \rangle := \frac{n_{\mathrm{v}}}{n_{\Box}}, 
\end{equation}
as the fraction of (detected) vortices $n_{\mathrm{v}}$ and lattice plaquettes $n_{\Box}$.
To characterize the BKT transition, we measure the spin-spin correlation $C_{\text{spin}}(|\textbf{r} - \textbf{r}'|)$ and the helicity modulus $\langle \Upsilon \rangle$~\cite{Berezinskii_1971, Sandvik_Proceedings_2010}.
The spin–spin correlation is denoted as
\begin{equation}
    C_{\text{spin}}(|\textbf{r} - \textbf{r}'|) = \langle \textbf{S}_{i} \textbf{S}_{j} \rangle = \langle \cos(\theta_{i} - \theta_{j}) \rangle
\end{equation}
considering all-to-all spin pairs $i, j \in [1, N]$ while excluding the trivial correlation $i = j$. 
The helicity modulus $\langle \Upsilon \rangle$, often referred to as the \textit{spin stiffness}, describes the response of ordered spins to a global twist caused by a perturbation~\cite{Sandvik_Proceedings_2010}.
Nelson and Kosterlitz have shown that the spin stiffness exhibits a universal jump at the transition temperature $T_{\text{BKT}}$ in the thermodynamic limit, known as the Nelson-Kosterlitz criterion~\cite{Nelson-Kosterlitz_1977}:
\begin{equation}\label{eq:NK-criterion}
    \lim_{L \rightarrow \infty} \langle \  \Upsilon(T_{\text{BKT}}) \ \rangle = \frac{2 \ T_{\text{BKT}}}{\pi} \ .
\end{equation}
Expanding the canonical free energy in the presence of a small perturbation $\phi$ gives
\begin{equation}
    \langle \Upsilon \rangle \equiv \frac{\partial^2 F(\phi)}{\partial \phi^2} \ ,
\end{equation}
see Ref.~\cite{Sandvik_Proceedings_2010} for greater detail.
We consider an x-directed perturbation $\phi$ and use a simplified notation in terms of the spin angle $\theta$. 
We follow the notation of de Andrade et al.~\cite{Andrade25}:
\begin{align}
\mathcal{H}_{x}                          &= \frac{1}{N} \sum_{\langle i, j \rangle_{x}} \cos(\theta_{i} - \theta_{j}) \\
I_{x}                           &= \frac{1}{N} \sum_{\langle i, j \rangle_{x}} \sin(\theta_{i} - \theta_{j}) \\
\label{eq:Upsilon-andrade}
\langle \Upsilon \rangle_{x}        &= \langle \mathcal{H}_{x} \rangle - N \beta  \langle I_{x}^{2} \rangle
\end{align}
where $\beta$ indicates the inverse temperature.
Our results provide qualitative evidence for a BKT transition in the simulated systems, characterized by the crossover from algebraic to exponential decay of the spin-spin correlation function, as shown in Fig.~\ref{supp: Correlation-Figure}a,b. 
We apply the Nelson--Kosterlitz criterion, see Eq.~\eqref{eq:NK-criterion}, to system sizes $L= 30, 40, 50, 60$, obtaining size-dependent estimates of the critical temperature. 
Fig.~\ref{supp: Helper-Figure}a illustrates the application of the criterion to the largest system ($L = 60$).
Combining these estimates with leading-order finite-size correction gives our final estimate $T_{c} = 0.903(7)$~\cite{Sandvik_Proceedings_2010, Hsieh2013, Andrade25}. 
This is consistent within uncertainties with the results reported by Gupta and Baillie~\cite{Gupta-Baillie}, $T_{c} = 0.894(5)$, and Schultka and Manousakis~\cite{Schultka_1994}, $T_{c} = 0.895(4)$. 
Our estimate lying slightly above is unsurprising, since both studies reached substantially larger systems (maximum sizes $L = 512$~\cite{Gupta-Baillie} and $L = 400$~\cite{Schultka_1994}).\\ 
As shown in Fig.~\ref{supp: Helper-Figure}b, the vortex density $\langle \rho \rangle$ exhibits the characteristic vortex proliferation of a BKT transition, supporting the reliability of the vortex-detection algorithm. 
The $U(1)$ correlation function is computed with open boundary conditions, restricting the analysis to vortices located within the central $L/2 \times L/2$ bulk region.
Based on these vortices, the proposed $U(1)$ correlation function indeed allows the BKT transition to be characterized with a quality comparable to that of the the spin-spin correlation function, as shown in Fig.~\ref{supp: Correlation-Figure}c,d.

\FloatBarrier
% BIBLIOGRAPHY
\bibliography{bib_T_star.bib}
\end{document}